\documentclass[%
reprint,
longbibliography,
 amsmath,amssymb,
 aps,
]{revtex4-2}

\usepackage{graphicx}
\usepackage{dcolumn}
\usepackage{bm}
\usepackage{braket}
\usepackage{wasysym}
\usepackage{mathtools}
\usepackage{color}
\usepackage{lineno}

\begin{document} 
\title{Chiral Selection, Isotopic Abundance Shifts, and Autocatalysis of Meteoritic Amino Acids} 

\author{Michael A. Famiano}
\email{michael.famiano@wmich.edu}
\affiliation{Dept. of Physics, Western Michigan University, Kalamazoo, MI 49008-5252 USA}
\altaffiliation{National Astronomical Observatory of Japan, 2-21-1 Mitaka, Tokyo 181-8588 Japan}
\altaffiliation{School of Physics and Nuclear Energy Engineering, Beihang Univ. (Beijing Univ. of Aeronautics and Astronautics),Beijing 100191, P.R. China}
\altaffiliation{Joint Institute for Nuclear Astrophysics}

\author{Richard N. Boyd}
\email{richard11boyde@comcast.net}
\affiliation{Dept. of Physics, Dept. of Astronomy, The Ohio State University, Columbus, OH 43210 USA}

\author{Takashi Onaka}
\affiliation{Dept. of Physics, Meisei University, 2-1-1 Hodokubo, Hino, Tokyo 191-8506, Japan}
\altaffiliation{Dept. of Astronomy, Graduate School of Science, Univ. of Tokyo, 7-3-1 Hongo, Bunkyo-ku, Tokyo 113-0033 Japan }

\author{Toshitaka Kajino}
\affiliation{National Astronomical Observatory of Japan, 2-21-1 Mitaka, Tokyo 181-8588 Japan}
\altaffiliation{School of Physics and Nuclear Energy Engineering, Beihang Univ. (Beijing Univ. of Aeronautics and Astronautics),Beijing 100191, P.R. China}
\altaffiliation{Dept. of Physics, Graduate School of Science, Univ. of Tokyo, 7-3-1 Hongo, Bunkyo-ku, Tokyo 113-0033 Japan }

\date{\today}


\begin{abstract}
The discovery of amino acids in meteorites has presented two clues to the origin of their processing subsequent to their formation: a slight preference for left-handedness in some of them, and isotopic anomalies in some of their constituent atoms. Numerous models have been developed to explain these phenomena. In this article we present theoretical results from the Supernova Neutrino Amino Acid Processing (SNAAP) model, which uses electron anti-neutrinos and the magnetic fields from source objects such as supernovae or colliding neutron stars to selectively destroy one amino acid chirality, and the same anti-neutrinos to create isotopic anomalies. For plausible magnetic fields and electron anti-neutrino fluxes, non-zero, positive enantiomeric excesses, $ee$s, defined to be the relative left/right
asymmetry in an amino acid population, are reviewed for two amino acids,
and conditions are suggested that would produce $ee>0$ for all of the $\alpha$-amino acids. This is accomplished by electron anti-neutrinos produced by the cosmic object interacting with the $^{14}$N nuclei in the amino acids. The relatively high energy anti-neutrinos that produce the $ee$s would inevitably also produce isotopic anomalies. A nuclear reaction network was developed to describe the reactions resulting from them and the nuclides in the meteorites. At similar anti-neutrino fluxes, assumed recombination of the detritus from the anti-neutrino interactions is shown to produce appreciable isotopic anomalies in qualitative agreement with those observed for D/$^1$H and $^{15}$N/$^{14}$N. The isotopic anomalies for $^{13}$C/$^{12}$C are predicted to be small, as are also observed.
Autocatalysis, a chemical enrichment of a small imbalance, may be necessary for the SNAAP model, or for any model suggested so far, to produce the largest $ee$s observed in meteorites. Thus a new autocatalysis model is developed that includes both the spins of the nuclei and the chirality of the amino acids. Autocatalysis allows the constraints of the original SNAAP model to be relaxed, providing more flexibility in its application, and
increasing the probability of meteoroid survival in sites where amino acid processing is possible.  The SNAAP Model thus explains all of the observed phenomena related to processing of the amino acids in meteorites. These results have obvious implications for the origin of life on Earth.
\end{abstract}

\maketitle

\section{Introduction}
One of the enduring mysteries in science has been the means by which Earthly amino acids have become nearly exclusively left-handed. Although molecular chirality, or handedness, was discovered in the Nineteenth Century by Pasteur \cite{pasteur}, and the pure left-handedness of the amino acids was deduced subsequently, the origin of the amino acid chirality has remained a mystery. 
The long-standing explanation for creation of amino acids originated with measurements of the 1950s \cite{miller53,miller59}, which showed they could be produced in a spark discharge within the gaseous conditions thought to have existed on early Earth, though it should be noted that
the specific results of this experiment question the validity of the early atmosphere model
presumed~\cite{cleaves08,abelson66,engel1962petrologic}. Regardless, this scenario would have produced equal numbers of left- and right-handed amino acids, $i.e.$, they would be racemic, in contrast to what is found to be the case. The several suggested means of converting these racemic amino acids to homochirality, or single-handedness, via Earthly processes were discussed by Bonner \cite{bonner91}, who concluded they would be unlikely to do so. General discussions supporting this conclusion have also been provided by Mason \cite{mason84} and Barron \cite{barron08}.

\textit{However, it was concluded by Goldanskii and Kuzmin \cite{goldanskii89}, and by Bonner \cite{bonner91}, that amino acid homochirality is essential for the perpetuation of life.}


Thus it is tempting to assume that a slight chirality could have been introduced on Earth by meteorites, which have been found to contain amino acids \cite{kvenvolden70,bada83,cronin97,cronin98,glavin09,herd11},  and in many cases do exhibit a slight preference for left-handedness.  After arriving on Earth, that asymmetry could have been amplified by autocatalysis \cite{frank53,kondepudi85,goldanskii89}, an effect that has been demonstrated in the laboratory (\cite{klussman06,breslow06, mathew04,soai14}) to be capable of converting slightly enantiomeric mixtures, $i.e.$, those slightly favoring one handedness, to homochiral ones. This then shifts the scientific question to the origin of the slight left-handedness observed in meteoritic amino acids.

We define enantiomeric excess as 
\begin{equation}
ee = (N_L - N_D)/(N_L + N_D), 
\end{equation}
where N$_L$(N$_D$) is the number of left- (right-) handed molecules of each type in an ensemble. Thus nearly all of Earth's amino acids have an ${ee}$ = 1.0, $i.e.$, they are left-handed and homochiral. 

Cold chemistry seems able to produce the amino acids in outer space~\cite{hasegawa93,garrod06,furuya15}, but does not appear capable of producing them with a nonzero enantiomeric excess. (Here ``cold chemistry'' is very generally defined
in terms of astrochemical processes, particularly those taking place on interstellar dust or in meteoric interiors.  While the process may be cryogenic, some small 
heating may be involved, though temperatures are generally assumed to be $\sim$ 10 - 50 K.) Thus some other mechanism must be invoked to perform a chirality, or 
handedness, selection. Several models have been created to explain how these ${ee}$s might have developed. Perhaps the most frequently cited one is the Circularly 
Polarized Light (CPL) model, which utilizes ultraviolet CPL, produced by first scattering the light from an extremely hot star by interstellar dust to polarize it, then 
letting it process the amino acids. It was first suggested by Flores et al. \cite{flores77} and Norden~\cite{norden77}, and subsequently elaborated in detail by many 
groups \cite{bailey88,takano07,takahashi09,meierhenrich10,demarcellus11,mein14}. This model has the advantage that its chiral selectivity can be experimentally 
demonstrated with beams of polarized photons from an accelerator. However, achievement of even a small $ee$ requires the destruction of most of the preexisting amino 
acids of both chiralities. The CPL model can produce either positive or negative $ee$s.

A completely different model, the Magneto-Chiral Anisotropy (MCA) model, was proposed by Wagniere and Meier \cite{wagniere82}, explored experimentally by Rikken and Raupach \cite{rikken00}, and developed further by Barron \cite{barron00}. In this model, the interaction between photons from an intense light source, for example, a supernova, and molecules in a magnetic field, possibly from the 
supernova's nascent neutron star or from a companion neutron star, would produce a chirality-dependent destruction effect on the amino acids. The dielectric constant of a medium depends on $\mathbf{k}\cdot\mathbf{B}$, where $\mathbf{k}$ is the wavevector in the direction of travel of the incident light and $\mathbf{B}$ is an external magnetic field.  The dielectric constant is different when incident light travels in the same direction as the external magnetic field from what it is when it travels in the opposite direction. This effect has opposite signs for L- (left-handed) and D- (right-handed) enantiomers.  The net result is that one enantiomer absorbs more of the incident light than the other, and thus is preferentially destroyed.  Experimental studies on this effect have resulted in $ee$s on the order of 10$^{-4}$ for chiral molecules \cite{guijarro09}. The MCA model can also produce $ee$s of either sign.

Although there are other explanations for the origin of a preferred amino acid chirality in outer space, we believe that they are less well developed than the CPL or MCA models~\cite{meierhenrich08,guijarro09}, or the Supernova Neutrino Amino Acid Processing (SNAAP) model \cite{boyd10,famiano18b, boyd18a}, which also appears able to explain how amino acids achieved left-handed chirality in outer space.In this model,
meteoric amino acids in magnetic fields are selectively destroyed via weak 
interactions with
an external lepton flux.  The leptons (in this case, anti-neutrinos from a stellar source), interact at different rates with the spin-1 nitrogen nuclei bound in  
left-handed and right-handed amino acids because the net magnetization - hence spin - 
vectors are oriented differently with respect to the anti-neutrino spin vector.  
It is important to note that, while the original SNAAP model was developed to accommodate conditions that might be found surrounding
a core-collapse supernova, it has been generalized significantly to explore models in which amino acids in magnetic fields and/or lepton fluxes might be
modified.  While the term \textit{SNAAP model} is maintained, in this paper it is used to denote very general conditions in which amino acids may be modified via
interactions with leptons in magnetic fields.
The interaction rate between the anti-neutrinos and nitrogen nuclei is spin-dependent.  (See 
\S\ref{snaap_desc} for more details.)
Recent efforts using quantum molecular calculations have shown that this model does 
produce amino acid chirality with significant $ee$s, and they are positive for most of the
amino acids studied \cite{famiano18b}. 
Two possible viable astrophysical sites for this model were suggested \cite{boyd18} to be 
neutron-star-Wolf-Rayet-star binary systems and binary neutron-star mergers. These seem to
provide the necessary features to make the SNAAP model work and avoid problems associated
with an isolated supernova.  

However, meteoroids that exhibit amino acid enantiomeric excesses have been found to have 
another feature that must be associated with the processing they undergo after the amino 
acids are formed: enhancement of certain isotopes over their terrestrial values. The 
D/$^1$H ratio has been found to have a significant enhancement
with meteoric sample measurements varying between 0.1 to around 7 times larger than 
terrestrial values.  The $^{15}$N/$^{14}$N ratios are also found to be significant, with ratios between about 5\% to 35\% larger than terrestrial values. 
Those for $^{13}$C/$^{12}$C are much smaller with enhancements less than about 5\%
terrestrial values while some samples are even deficient in $^{13}$C relative to $^{12}$C
by about 1\%. The D/$^1$H enhancement has been treated as a result of the cold
chemistry that amino acids would undergo~\cite{hasegawa93,garrod06,furuya15}, but the
other enhancements have been found to be more difficult to explain by chemical means~
\cite[see discussion in][]{elsila12,pignatari15}. 

The $ee$s and isotopic enhancements have generally been dealt with separately. However, we show in this paper that they may have a common cause or trigger. Here, we examine
the results of changing the total electron anti-neutrino fluence on the
initial enantiomeric excess and the isotopic abundance ratios.

In Section II, we review the basic features of the SNAAP model. 
Section III discusses the autocatalysis models used in evaluating final enantiomeric excesses, and we show that even low-field, low-flux 
scenarios may result in substantial $ee$s given enough chemical processing time.  Section IV presents the results of the calculations of the isotopic anomalies. Section V presents our conclusions.


\section{The SNAAP Model}
\label{snaap_desc}
The SNAAP model has been described in multiple prior publications
\cite{boyd18,boyd18a,famiano18,famiano18b,famiano19}, and will be 
summarized here with some prior results.  Since the purpose of this paper is to
explore how the SNAAP model can be enhanced with an autocatalytic mechanism and not the
original SNAAP mechanism, the reader
is referred to prior references.

In the scenario described by this model, meteoroids in external 
magnetic fields can be processed in an anti-neutrino flux from one of
several objects.  
The nuclear physics that describes the weak interactions between the incident (inherently chiral) anti-neutrinos and
the $^{14}$N nuclei within amino acids are spin dependent, necessitating inclusion of spins in the SNAAP model. 
The external magnetic field allows for a non-zero magnetization (and net spin orientation) of
the $^{14}$N nuclei, and selective destruction of some nuclei via the reaction
\begin{equation}
\bar\nu_e + ^{14}N \rightarrow e^+ + ^{14}C
\end{equation}
takes place.  
The matrix element that couples the initial state of $\bar\nu_e$, the anti-electron neutrino, and $^{14}$N with the final state of e$^+$, an anti-electron, or positron, and $^{14}$C is
\begin{equation}
\braket{\bar\nu_{e}  {^{14}N} | T | e^{+}  {^{14}C}},
\end{equation}
where T is the weak interaction operator, which is the same as that for nuclear $\beta$-decay.

Conservation of angular momentum and parity from the spin-1 $^{14}$N (positive parity) ground state to the spin 0 (positive parity) ground state of $^{14}$C requires two units of angular momentum to come from the $\bar\nu_e$ or the positron wave functions.
This is known from basic nuclear physics \cite{boyd08} to produce roughly a two order of magnitude smaller cross section when the $^{14}N$ and $\bar\nu_e$ spins are aligned  than when they are anti-aligned. 
The net result is an approximate interaction cross section $\sigma$ that varies with the angle $\phi$ between the direction of the $^{14}$N spin and the spin of the anti-neutrino as
\begin{equation}
\sigma \propto 1 - cos\phi
\end{equation}
Note that this simple relationship exists only because $^{14}$N is a spin 1 nucleus.
This is the origin of the SNAAP model's ability to produce  $ee$s in amino acids.

Other spin-selective nuclear destruction reactions may also take place.  However,
the $^{14}$N destruction reaction dominates owing to its low Q-value and low transition order.

It has been shown previously \cite{famiano18,famiano18b}, that
the nuclear spin can be coupled to the molecular chirality in external
fields through the asymmetric part of the magnetic shielding
tensor \cite{buckingham04,buckingham06}.  Thus, the interaction 
of $^{14}$N with the $\bar\nu_e$ is sensitive to the molecular
chirality.  This effect is very small, about 1 part in 10$^6$.  However,
thermal equilibrium and low neutrino interaction rates allow
for an increased gradual reduction in the D-enantiomer over the 
L-enantiomer.  It will be shown here that this effect can be coupled to
an auto-catalytic mechanism in which a source of amino acid creation
can enhance this effect - in some cases to a homochiral state.

The destruction mechanism is nuclear, but amino acid chirality is molecular, so it must be shown how the nucleus and molecule are coupled. The external magnetic field aligns the $^{14}$N nuclei via their nuclear magnetic moments, whereas the effective electric field aligns the molecular electric dipole moments, which depend on the molecular chirality. This modifies the magnetic field at the nucleus by the effects of the orbital electrons, known as shielding - a phenomenon central to nuclear magnetic resonance~\cite{famiano18}.  
Because the shielding tensor depends on the electronic orbital configuration, the off-diagonal elements are sensitive to a parity transformation, and thus can change sign with chirality.

In an isotropic medium, the molecules tumble freely, and the average over the shielding tensor elements in a magnetic field is quite small. However, for molecules polarized in an external electric field, the temperature-dependent transverse shift, $\Delta_T\mathbf{B}$ can be written (in the linear approximation) as~\cite{buckingham06}:
\begin{equation}
\Delta_T\textbf{B}^{(N)} = (1/6kT) \varepsilon_{\alpha\beta\gamma} \sigma_{\alpha\beta}\mu_\gamma\textbf{B}^{(0) } \times \textbf{E}
\end{equation}
where $\varepsilon$ is the permutation operator and $\mu_{\gamma}$ is the molecular electric dipole moment in the $\gamma$ Cartesian direction. This shift is of opposite sign for left-handed and right-handed molecules. The shift is also perpendicular to the external $\textbf{B}$ and $\textbf{E}$ fields. For meteoroids in motion with respect to a highly magnetized body, the Lorentz force electric field will be created in the molecular rest frame. Thus the molecules will be polarized, and the sign of $\Delta \textbf{B}^{(N)}$ for nuclei within the molecule will depend on the molecular chirality.
This in turn causes a chirality sensitive perturbation on the magnetic orientation of the nuclei, which leads to a chirality-dependent magnetization (a bulk property).

\begin{figure}
    \centering
    \includegraphics[width=\linewidth]{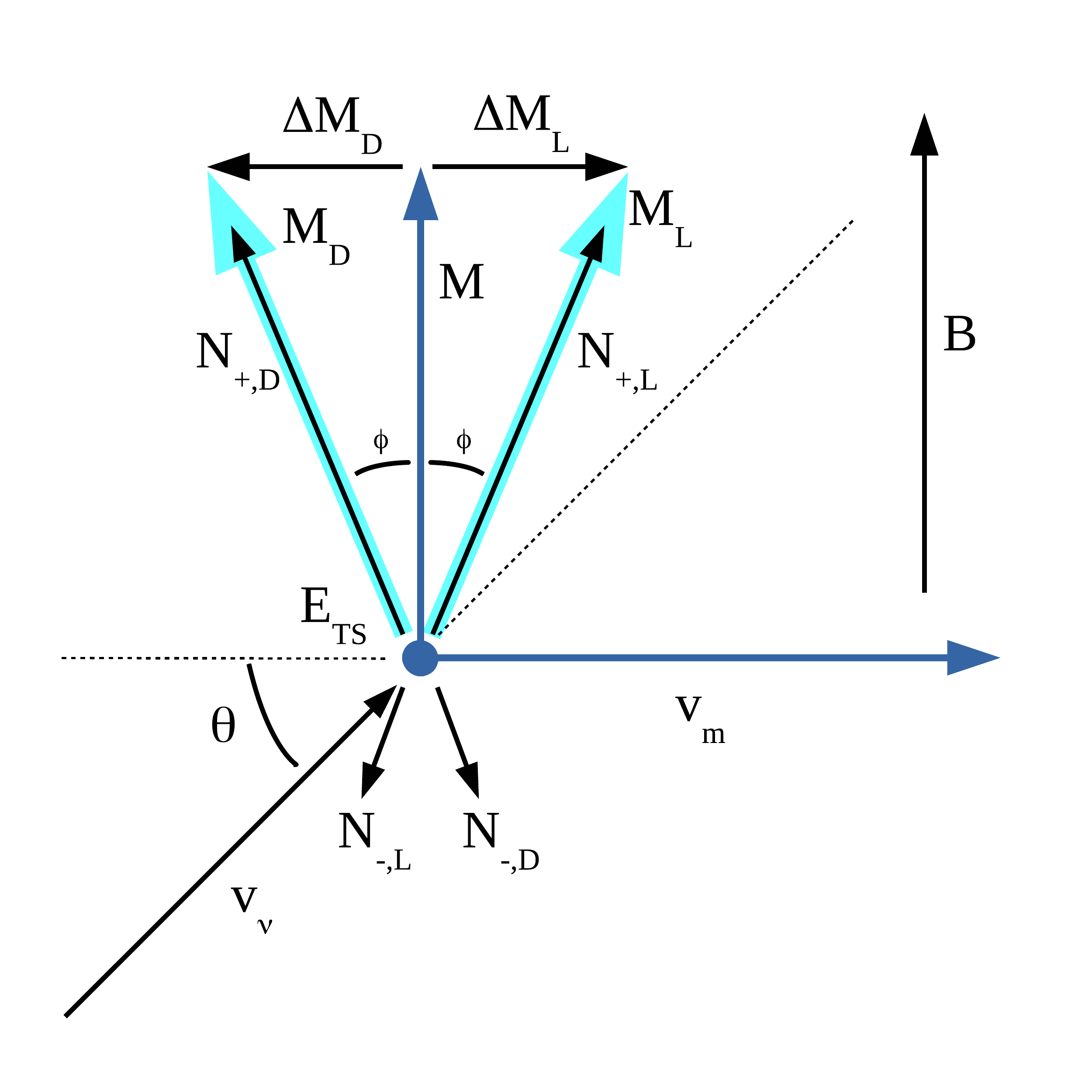}
    \caption{\label{fig:vectors}
    The vectors relevant to the processing of amino acids in this model. The vectors and labels are explained in the text~\cite{famiano18}.  Used with permission of Astrobiology.
    }
\end{figure}

The vectors associated with this scenario are illustrated in Figure \ref{fig:vectors}~\cite{famiano18}.
There it can be seen that the external electric field vector $\mathbf{E}_{TS}$ (which is 
coming out of the page in this diagram) is induced by the meteoroid's velocity vector 
$\mathbf{v}_m$ through the external magnetic field $\mathbf{B}$.  Here, an anti-neutrino 
velocity vector $\mathbf{v}_{\bar{\nu}}$ makes an angle $\theta$ with respect to the 
meteoroid velocity vector. The bulk magnetization vector is $\mathbf{M}$ for a meteoroid at
rest.  For moving meteoroids, the induced electric field creates additional transverse
magnetization components $\Delta\mathbf{M}_\chi$ where $\chi$ represents the chiral state.
This induced magnetization is chirality-dependent, and results in net positive and negative
spin components aligned along the  magnetization vectors.  The population of nuclei with
spins along these components are labeled as $N_{+,-}$ in the figure.  The angle 2$\phi$ is
the separation of the net magnetization vectors $\mathbf{M}_\chi$, where $\phi
=\tan^{-1}(\Delta M/M)$.  The difference in angle between the net magnetization and the
neutrino velocity results in different reaction rates for the different 
chiral states .

These components exist even without the coupling to the electric dipole moment \cite{buckingham06}, but that coupling enhances the difference between the angles that the two chiral states make with the anti-neutrino spin, hence of the chirality selective destruction of the amino acids \cite{famiano18b}. From the magnitude of these effects, one can determine the expected $ee$s for amino acids from the SNAAP model.

\subsection{Considerations From the Merger of Two Neutron Stars}
Two neutron stars in close orbit may have begun their existence as two massive stars. When the first exploded as a supernova, it became a neutron star that drew the outer one or two shells from the remaining massive star, creating an accretion disk around the neutron star, leaving the other as a Wolf-Rayet star. 

The recently observed merger of two neutron stars (NSs) \cite{abbott17}, event GW170817, demonstrated several important things. First, it showed that such events do occur and, unless human beings were incredibly lucky, are not entirely rare. Second, it reaffirmed the capability of gravitational wave detectors to produce 
forefront science. Third, it confirmed the predictions of general relativity. Finally, its observation in gamma-rays by the FERMI \cite{goldstein17} and INTEGRAL \cite{savchenko17} gamma ray detectors, and by the CHANDRA \cite{troja17} x-ray detector, as well as by many optical telescopes, determined that a lot of heavy nuclides were
synthesized via the rapid-neutron-capture-process (or r-process) resulting from two NS mergers. While the details of this depend on the masses of the two stars prior to merger, the final state neutron
star or black hole may not be of great consequence to our considerations. The actual amount of heavy nuclides created has been estimated \cite{kasen17} to be several tenths of a solar mass. Since this was made largely from what was initially neutron matter, a huge flux of electron anti-neutrinos must have been produced to convert these extremely neutron-rich progenitors to the r-process nuclei. 

\textit{But the two NS merger may also have served as an incubator for creating the molecules of life.}

Fortunately enough theoretical work \cite{rosswog03,perego14} has been done on two NS mergers that good estimates of the parameters needed to perform SNAAP model calculations exist. In particular, we know that the maximum magnetic field generated at the composite neutron star is around 10$^{17}$ G \cite{rosswog03}. That permits the magnetic field orientation region to extend nearly an order of magnitude beyond what it would be for the supernova from a neutron-star-WR-star binary system, another potential site for the SNAAP model. 


Perego et al. \cite{perego14} and Rosswog and Liebend{\"o}rfer \cite{rosswog03} calculated both the expected fluxes for electron neutrinos and anti-neutrinos, and their energies. The electron anti-neutrinos are the dominant species, and their total flux is expected to exceed 10$^{53}$ ergs in the fraction of a second during which they would be emitted. 

An important consideration is the region in which amino acids might be formed in the accretion disk around one of the neutron stars. Presumably the cooler outer regions can permit the creation of the complex molecules that enable the synthesis of amino acids, and may be shielded from the radiation from the WR star, for the portions of their trajectory that are closest to the WR star, by the more inward regions of the disk. Most disk simulations do not extend to temperatures at which molecules might form, but they agree that the temperature falls off roughly as r$^{-3/4}$. D'Alessio et al. \cite{dalessio01} found that the midplane temperature of a disk depends on its assumed size, but was typically several hundred K at 1 AU for the system they considered. Amino acids have been shown to form on dust grains \cite{munozcaro02}, although under different conditions than would be expected in the outer regions of an accretion disk. None the less, the conditions required would be expected to exist also in the disk. Thus molecules could begin to form around an AU from the central object \cite{hasegawa93}. 

Although there may be considerations that would allow amino acids to exist closer to their parent neutron star, for example, settling and mixing within the disk, this might be a long-term scenario which would result in the destruction of the amino acids formed in all but the largest meteoroids. Thus it might be difficult to have very many amino acids existing within 0.1 AU, the distance from a single neutron star that is necessary for them to experience a sufficiently large magnetic field to sustain an adequate selection between chiral states for an appreciable $ee$ to develop.

However, this would not necessarily be the situation for two NS mergers. When the second 
neutron star began to orbit the first at a distance where it would begin to intercept the 
outer regions of the disk, the disk would be disrupted by the gravitational field of the 
second star. Amino acid laden meteoroids might be pulled into a close orbit with the second
star, or might be deflected into elongated orbits. As the second star continued to plow
through increasingly dense regions of the disk, it would thoroughly mix the disk material, 
dragging some of the meteoroids that had formed in the disk
from the outer disk regions into regions closer to one or 
the other neutron star. As the two stars grew even closer, this disk material would begin 
to orbit both stars, and would be compressed under the increased gravitational potential of
the two stars. Meteoroids that had been deflected into highly elliptical orbits would
occasionally recur at close distances to one or both stars but, if they were large enough,
would not remain nearby long enough for their amino acids to be destroyed. Indeed, these
objects might represent the best candidates for processing their amino acid containing
meteoroids if they happened to be close to the two NSs as their burst of anti-neutrinos
occurred.

The observations of 21/Borosov \cite{guzik19} and Oumuamua \cite{meech17}, both thought to
be comets that are not from our Solar system, suggest that such comets are not rare.
Presumably the frequency of such objects in the vicinity of colliding neutron stars,
resulting from the destruction of at least one accompanying accretion disk, would be much
higher before they had time to disperse.

The huge magnetic field of the resulting neutron star, along with the enormous flux of
electron anti-neutrinos produced at merger, would surely produce significant $ee$s in many
of the amino acids contained in nearby meteoroids. The expanding “butterfly” inner disk of
matter that is predicted to occur \cite{rosswog03}, would presumably eventually push the
objects lying further from the center of the stars into outer space, there to seed the
surrounding volume with r-process nuclei and enantiomeric amino acids.
Could a two NS merger seed the Galaxy with enantiomeric amino acids? There was speculation
that the heavy nuclides produced by GW170817 could provide a significant fraction of those
observed in the Galaxy, and if that is the case, it might also have produced amino acids
with $ee$s. Although the distribution of enantiomeric amino acids need not be uniform
throughout the Galaxy, the recent observation \cite{ko19} of a huge overabundance of
$^{60}$Fe in Earthly deep ice core samples suggests that a massive cosmic event, possibly a
supernova or even two NS merger, did occur the order of a million years ago near the Solar
System. Furthermore, the continuing arrival on Earth of meteorites bearing enantiomeric
amino acids suggests that neither the anomalous $^{60}$Fe nor the enantiomeric amino acids
is a rarity.

We note that recent analyses have estimated the current rate of neutron star
merger events in the Galaxy to vary between  $\sim$1.5$\times$10$^{-6}$ -- 2$\times$10$^{-5}$ 
yr$^{-1}$~\cite{voss03,merger1,merger2,merger3}.  Given the amount of time for NS binaries to form
and undergo orbital decay, this number has not been constant.  However,
it is estimated to have reached a rate of about 10$^{-6}$ 
yr$^{-1}$ after about 2 Gyr.  If we roughly assume a merger rate of 10$^{-6}$ 
yr$^{-1}$ for the last 10 Gyr, then this would be a total of 10$^4$ mergers
in the Galaxy.  While this rough estimate a number of
possible events in which meteoric material might be processed, there are still many variables which must be accounted for,
such as the formation of a nearby planetary system within a sufficient time
window.

\subsection{Enantiomeric Excesses}
Calculations have been performed of the level of enantiomerism of amino acids that might have existed close to the center of the two NS system as they merged. Important factors include the gravitational field, the magnetic field, the meteoroid orbital characteristics, and the electron anti-neutrino flux.
Many of these factors depend on the meteoroid distance from the neutron stars. This includes the meteoroid velocity, which is closely linked to the distance from the stars.  The details of the 
calculation, including the sensitivity of the resultant $ee$s to various input parameters, are described in prior work \cite{famiano18}, so will only be summarized here. 

The shielding tensors and electric dipole moments were computed for 
each enantiomer of the $\alpha$-amino acids using the \texttt{Gaussian16} \cite{g16} quantum chemistry code \cite{famiano18b}. Calculations included amino acid ligands,
zwitterions, and ions in water solution.
The environmental conditions which approximate the expected environment in the space surrounding a two NS merger were simulated. For a typical merger event, two stars of 1 solar mass each with a net surface field of 10$^{11}$ T were assumed.  Because the dynamics of the fields (gravitational and magnetic) and the anti-neutrino flux can be complicated in a typical event, we have assumed a spherical mass of 2 M$_\odot$ with a dipole field. 

A constant anti-neutrino flux of 10$^{57}$ cm$^{-2}$s$^{-1}$ was assumed at the surface of the merger 
event for one second with an average peak cross section (for anti-aligned $\bar\nu_e$ and $^{14}$N) for the $^{14}$N($\bar{\nu}_e,{e^+}$)${^{14}}$C reaction of 10$^{-40}$cm$^2$.  The net anti-neutrino interaction rate, $f$, relative to half the 
$^{14}$N relaxation time, $T_1$ 
($= \lambda_R^{-1}$ and $\lambda_{\bar{\nu}}\sim 10^{-42} - 10^{-40}$), is defined by a unitless fraction \cite{famiano18}:
\begin{equation}
f \equiv 2\frac{\lambda_{\bar{\nu}}}{\lambda_R} = 2\lambda_{\bar{\nu}}T_1
\end{equation}

Using the computed shielding tensor, the nuclear magnetic polarizability for the cationic form of isovaline was computed with a density functional theory (DFT) calculation using a \texttt{pcS-2} 
basis set
Prior to this, the initial electronic wavefunctions were optimized using an MP2
computation with the \texttt{aug-cc-pVDZ} basis. NMR properties, including the shielding tensors,
were then computed.
With the resultant shielding tensor asymmetries, the difference in magnetic field vector for each
enantiomer at the $^{14}$N nucleus
$\mathbf{B}_{\chi} = \mathbf{B}_\circ+\Delta\mathbf{B}_\chi$ was determined, where $\chi$ represents the chirality
of a particular enantiomer. The anti-neutrino interaction rates vary as
$\mathbf{\sigma}\cdot\mathbf{B}_\chi\rightarrow\mathbf{v}\cdot\mathbf{B}_\chi$ \cite{morita}, where $\mathbf{\sigma}$ and $\mathbf{v}$ are the anti-neutrino spin and velocity vectors respectively.
Assuming a massless anti-neutrino, its spin vector points in the same direction as its momentum
vector.  
For a massive Dirac neutrino with a mass limit of $\sim$1 eV~\cite{katrin}, the particle's helicity and chirality are no longer synonymous, and a boost induced by the meteoroid's reference frame could
result in a flip in the anti-neutrino helicity.  However, for a 10 MeV anti-neutrino, this would require a boost equivalent to $\gamma\approx$2000, which is nearly the speed of light and does not occur in
this model.  Alternatively, a Dirac mass may result in chiral mixing for a particular state.
For a 10 MeV neutrino, chirality mixing would result in a very small change in the overall population of 
left-handed and right-handed neutrinos - on the order of 1 part in 10$^7$~\cite{griffiths08,perkins_2000}. 
Similarly, neutrino Majorana mass could result in chirality mixing without violating local symmetries~\cite{balantekin18}.  
However, right(left)-handed (anti)-neutrinos may not interact via the parity violating weak force~\cite{wong}.  Given that all currently 
observed neutrinos are ultra-relativistic, massive neutrinos are not expected to contribute significantly to 
the results of this model, and would only be a null term in the results as opposed to an opposing term.

\subsection{Results of \textit{ee} Calculations}
Several simulations were run in previous works~\cite{famiano18,famiano18b,boyd18,boyd18a} in which the $ee$s of isovaline and alanine 
were computed as a function of time/anti-neutrino exposure in the vicinity of a two NS merger. We review some results here.  While prior results used a linear approximation
of the molecular polarization, here, we use an exact molecular polarization in 
the induced electric field.

\begin{figure}
    \centering
    \includegraphics[width=0.490\textwidth]{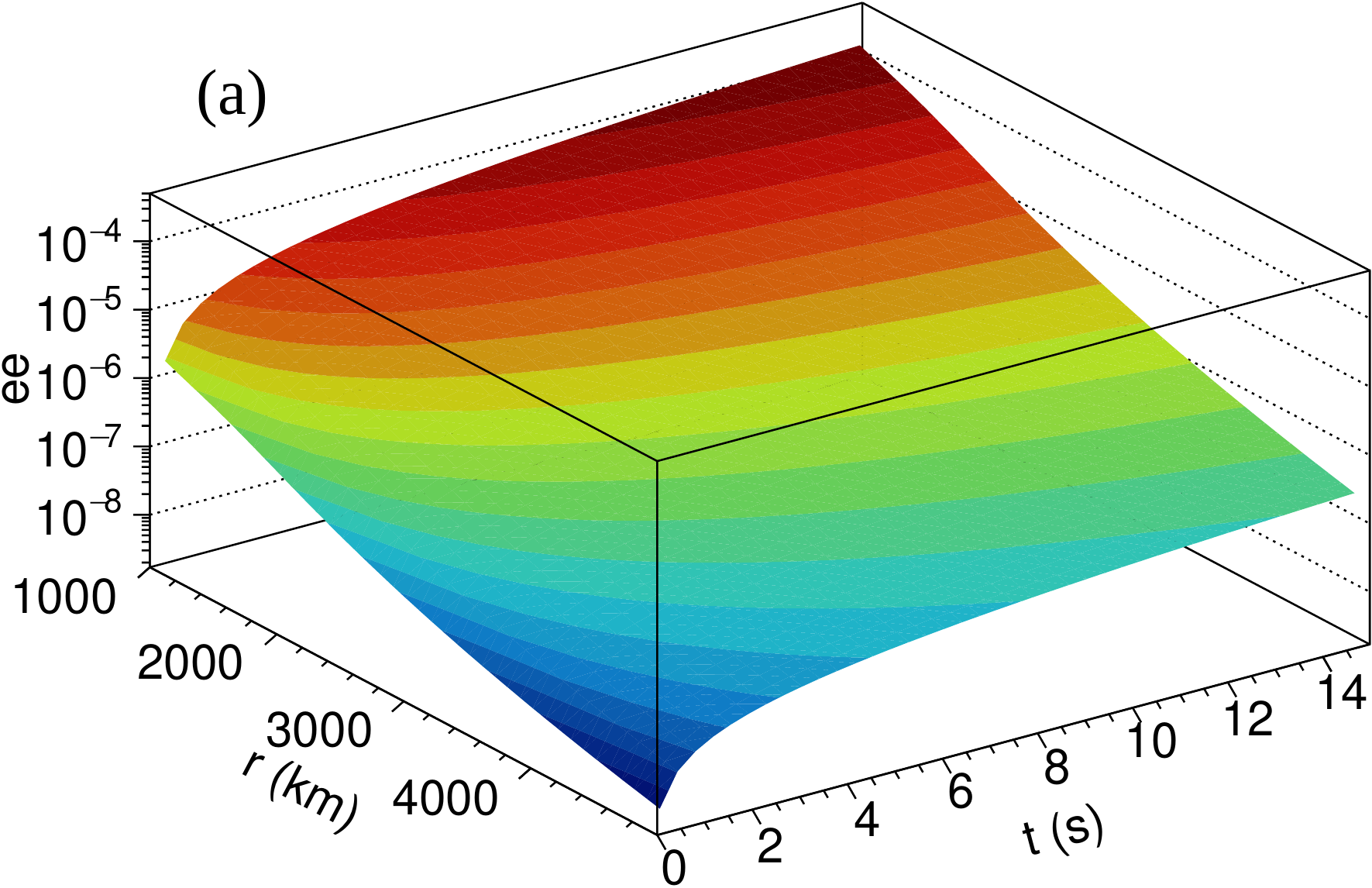}
        \includegraphics[width=0.49\textwidth]{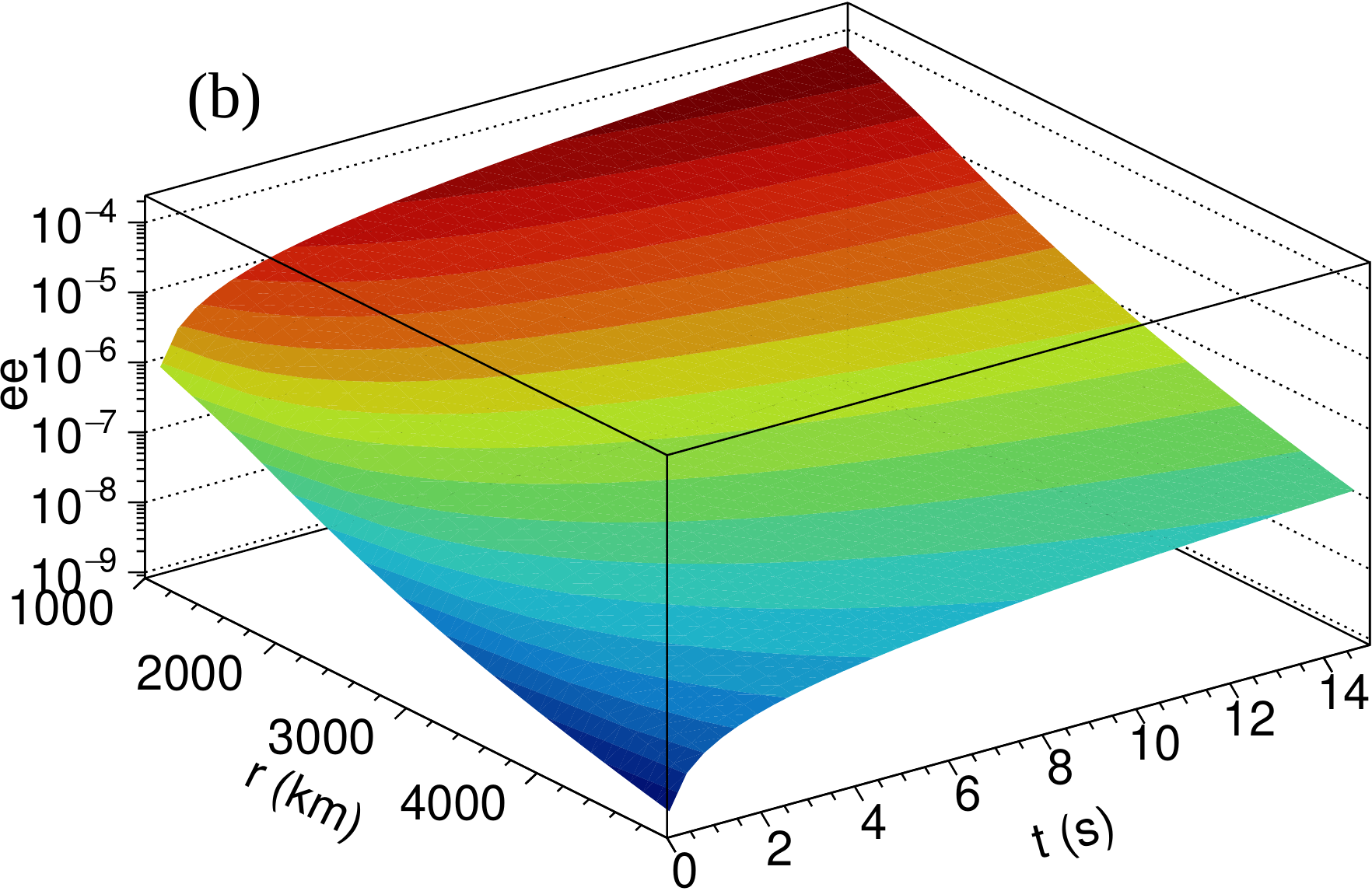}
    \caption{
        The $ee$ for cationic isovaline (a) and zwitterionic alanine (b) as a function of time and radius from a two NS merger. The velocity was assumed to be 5\% of the orbital velocity, as is explained in the text.
        }
    \label{fig:ees}
\end{figure}

The net $ee$s as a function of time and distance of approach at the time of anti-neutrino exposure for cationic isovaline and zwitterionic alanine are shown in Figure \ref{fig:ees}, and that for alanine in Figure \ref{fig:ees}.
Because the anti-neutrino burst is short, the order of a second, it can 
be assumed that the separation radius changes little over the course of 
an event. Disk viscosity, however, is an unknown parameter, so a constant
meteoroid velocity of 5\% of the vacuum orbital velocity was assumed.  
While this might appear to be so low a velocity that the meteoroid would fall into the central object, it was chosen to produce a realistic result for $ee$s, not to have anything to do with orbital dynamics. This velocity was chosen to accommodate the fact that the velocity and the magnetic field will rarely be perpendicular to each other. Even at this small velocity the effective electric field is so large that the molecular polarization is essentially 100\%.
For larger electric fields the linear response we have assumed becomes invalid.

It can be seen from Figure \ref{fig:ees} that the $ee$s achieved for isovaline
and alanine
can become large, up to 0.01 percent. This is well below the $ee$s measured 
in meteoric analyses, at least for the range of values of radius assumed. Smaller
separation distance would produce larger $ee$s, but they would still be well below the few percent observed for amino acids in some meteorites.

Note that this factor and that for isovaline are about two orders of magnitude less than reported in previous publications \cite{famiano18, famiano18b}. This is due to our discovery that the approximation used routinely in calculating the matrix elements relevant to strong terrestrial magnetic fields fails at the extreme fields of cosmic objects. Solving the equations exactly is responsible for the reduction in $ee$s.
Although the present results are presented for the isovaline cation and zwitterionic alanine,
our previous
work~\cite{famiano18b} strongly suggests that similar results would be
obtained for most other amino acids.

The conditions in Figure \ref{fig:ees} are a direct
result of the large magnetic fields, 
which together with meteoroid motion
produce large local electric fields. These may exceed the dielectric
strengths of the amino acids.  However, this effect could be mitigated in
several ways.  A stronger magnetic field and a weaker electric field can
result in the same chiral selection. Additionally, this effect could also
work if the amino acids are contained in crystalline structures, for
which a much smaller electric field would suffice for $ee$ production.

In prior work, it was shown that, if the anti-neutrino flux is too
intense or continues for too long, a total destruction of both L- and
D-enantiomers can result, and a sudden drop in $ee$s occur as all the
amino acids are destroyed. This is partially due to their thermalization.
In the case of the two NS merger event, the anti-neutrino pulse
is so short that the $ee$s increase rapidly and the anti-neutrino flux
stops well before all of the amino acids are destroyed.  

\subsection{Variations in Isotopic Abundance Ratios}
Because the model presented here occurs as a result of weak interactions in
nuclei bound in amino acids, the nuclear abundances will also change
as a result of these interactions.  The produced nuclei can then go
on to form other amino acids or biomolecules within the mixture.  These molecules will contain the nuclear resulting from the prior processing.  Additionally,
the formation of D, $^{15}$N, and $^{13}$C from the capture of neutrons 
liberated in the neutrino capture process may be exhibited in
the abundance ratios of biomolecules contained within the meteoroid.
This effect was studied in previous work~\cite{famiano19} and will be summarized here.

While the SNAAP model naturally results in shifts in isotopic abundances of the 
involved nuclear species, it can operate in addition to existing mechanisms without replacing those
mechanisms.  For example, mechanisms in which isotopic enrichment may have occurred prior to processing within
meteorites may complement the existing model.  In fact, these mechanisms operating alongside the SNAAP
model may be responsible for the large variations in isotopic abundance ratios in many 
meteoritic molecules.  For example, deuterium and $^{15}$N enrichment may have occurred prior to chemical processes
and the resulting enantiomeric excesses may be a reflection of this enrichment~\cite{caselli12,ceccarelli14}.  Thus these 
processes are not to be ruled out. They may provide complementary mechanisms by which isotopic fractionation
may occur in meteoritic amino acids.  Changes in isotopic abundances by the weak interactions in the SNAAP model 
may provide the initial isotopic ratios found in the products of subsequent chemical processes.  
While the SNAAP model may produce small initial enantiomeric excesses, it may be the initial 
mechanism in a sequence including subsequent astrochemical mechanisms
responsible for production of much larger $ee$s in meteoritic environments.

Neutral-current and charged-current reactions 
were studied in a nuclear reaction network including nuclei up to the iron
region.  Neutrino and anti-neutrino induced
reactions assuming electron and $\mu$ neutrinos were included.  Neutrino captures, proton spallation, 
neutron spallation, and alpha spallation reactions were included.  Subsequent
$\beta$ decays were followed in the reaction network as well.  The produced neutrons were tracked in subsequent post-processing, and
neutron captures on existing nuclei were also included.
Cross section data sources are discussed in a prior publication~\cite{famiano19}.  For the neutron captures, the 
	experimental uncertainties are typically quite low.  For example, uncertainties for thermal neutron captures on deuterium are 
	$\approx$3\%~\cite{endf}.  For $^{13}$C neutron captures, the uncertainties are $\sim$2\%~\cite{daub13}.  Similar uncertainties 
	surround the remaining neutron cross sections in this region.  The neutrino interactions are less certain.  For example,
	neutral-current cross sections on $^{12}$C, for which no change in charge occurs for the target nucleus (e.g., elastic scattering)  
	can be as large as $\sim$30\%~\cite{suzuki06} for the shell model used in this project.  Decay
	lifetimes
	are taken from experimental measurements and are accurate to well under a fraction of 1\%.   The neutrino capture uncertainty 
	is certainly the largest contributor to the overall uncertainty and can result in an overall uncertainty of the total number of 
	neutrons produced. It was shown in a previous paper that neutron captures are responsible for nearly all of the 
	D/H change, about 90\% of the $^{15}$N/$^{14}$N change, and only about 10\% of the $^{13}$C/$^{12}$C change.  Neutrons are produced
	almost exclusively via neutrino interactions in the surrounding rock (e.g., interactions in iron).  Given this, we can conservatively
	estimate the overall uncertainties in $\delta$D at 30\% and in $\delta^{15}$N at about 27\%. Note that the isotopes corresponding to 
	the denominators in these ratios
	are so abundant that neutrino rates and neutron captures on these rich isotopes has a negligible effect.

From this reaction sequence, the relative nuclear abundances were
evaluated in the biomolecular mix within the 
meteoroid, where the isotopic deviation, $\delta^AZ$, of
a specific isotopic species of charge, Z, and mass, A, is defined as:
\begin{equation}
    \delta ^AZ\equiv \left[\frac{\left(\frac{Y_A}{Y_{A_0}}\right)_m}
    {\left(\frac{Y_A}{Y_{A_0}}\right)_0} - 1\right] \times 1000\permil
\end{equation}
where $(Y_A/Y_{A_0})_m$ is the ratio of the abundance of an isotopic species of mass $A$ to 
that of the most abundant isotope of that element in the meteorite, denoted by the subscript $m$. That quantity is divided by the same ratio in an isotopic 
standard (indicated by the subscript $0$).  

In this case, the isotopic standard for $^{15}$N/$^{14}$N is Earth air ratios~\cite{junk58}, while that for D/H is 
the Vienna Standard Mean Ocean Water (VSMOW) standard~\cite{hagemann70,dewit80}, and $^{13}$C/$^{12}$C is the Vienna PDB (VPDB) standard~\cite{hut87}.  All of these are taken from terrestrial samples.

For the present calculations, the reaction network was run for up to 
10 s, at which time 
the total integrated anti-neutrino flux would be equal to the maximum
expected from the two NS merger \cite{rosswog03}. The value of
$\delta$, the isotopic enrichment, was computed at each time step in the evaluation. 
The full time
corresponds to the electron anti-neutrino flux estimated~\cite{rosswog03}
at a distance of 10,000 km from the center of the two NS merger source.

The dependence on initial isotopic abundances was also explored in this
evaluation.
One assumed abundance set 
corresponded to the CI abundances of the Orgueil meteorite rock average
as stated in Lodders et al.~\cite{lodders09} with isotopic distributions
for individual elements taken to be that of the solar system 4.6 billion
years ago (Table 10 from the same reference). Additionally, two other
abundance distributions were chosen to match the abundance distribution
of the solar system at formation~\cite{lodders09} and the current solar
photospheric values~\cite{lodders09}.

An important feature involves the effects of the neutrons as they
thermalize and are finally captured. Their thermalization occurs mostly 
on protons, either bound to amino acids or existing in other molecules, 
especially water. Their initial energy of around 200 keV is 
certainly enough to dislodge a target proton from an amino acid or other 
molecule, so some destruction of amino acids and other molecules may occur following the neutrino and anti-neutrino interactions prior to neutron thermalization. However, the capture of a 
thermal neutron on a proton occurs at much lower energy, so that might 
allow conversion of a proton in an amino acid to a deuteron without 
disrupting the molecule. The same considerations would apply to 
conversion of H to D in other molecules.

Another potential complication involves the molecular destruction that will result from the recoiling positrons and electrons produced in the neutrino and anti-neutrino induced reactions. These might be produced with several MeV of energy, and will produce copious molecular destruction as they lose their energy via interactions with molecular electrons. Furthermore, the first few $e^+ e^-$ interactions will result in electrons with sufficient energy to do additional molecular destruction as they lose their energy. However the resulting atoms and molecular fragments will not have additional isotopic changes, rather they will simply contribute to the atomic and molecular fragment aggregate that exists immediately following the neutrino and anti-neutrino burst. 

Given the difficulty of estimating the results of either of these two effects, they were not
included in our estimates of the isotopic anomalies. None the less, they could be important,
and may ultimately require more study if stricter comparisons to data are warranted.

Isotopic abundances were calculated using the three different two NS merger scenarios of neutrino and anti-neutrino emission 
from Rosswog and Liebend{\"o}rfer~\cite{rosswog03}, and with different 
assumptions of the initial abundances in the meteoroids. Although the neutrino/anti-neutrino scenarios describe the emission of all such species, the electron anti-neutrinos dominate, so the following discussion will refer just to `anti-neutrino' emission, with the others, most notably electron neutrinos, implied. The effects of 
these different scenarios are shown in several figures in Famiano et 
al.~\cite{famiano19}. 
\begin{figure}
    \centering
    \includegraphics[width=0.49\textwidth]{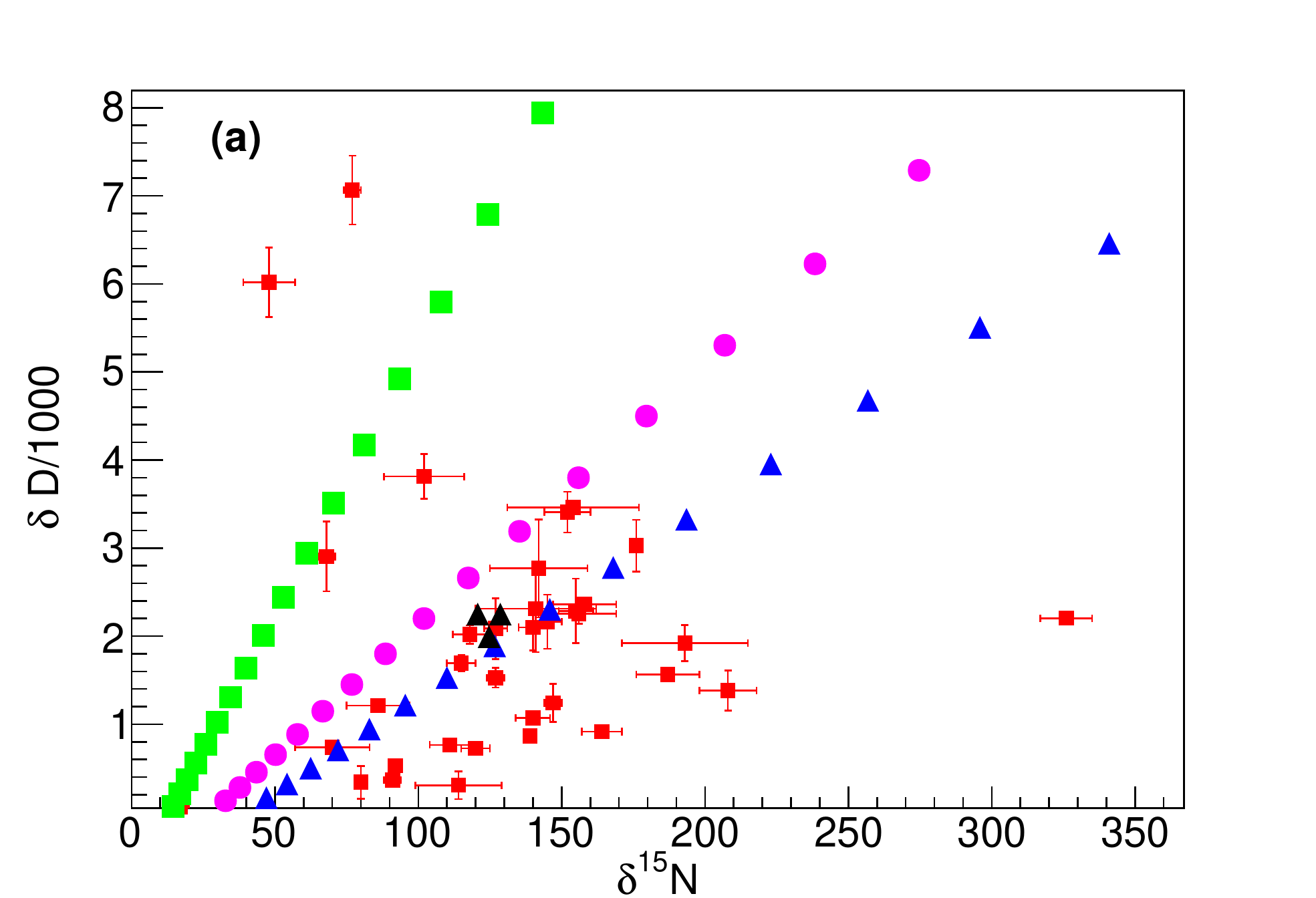}
     \includegraphics[width=0.49\textwidth]{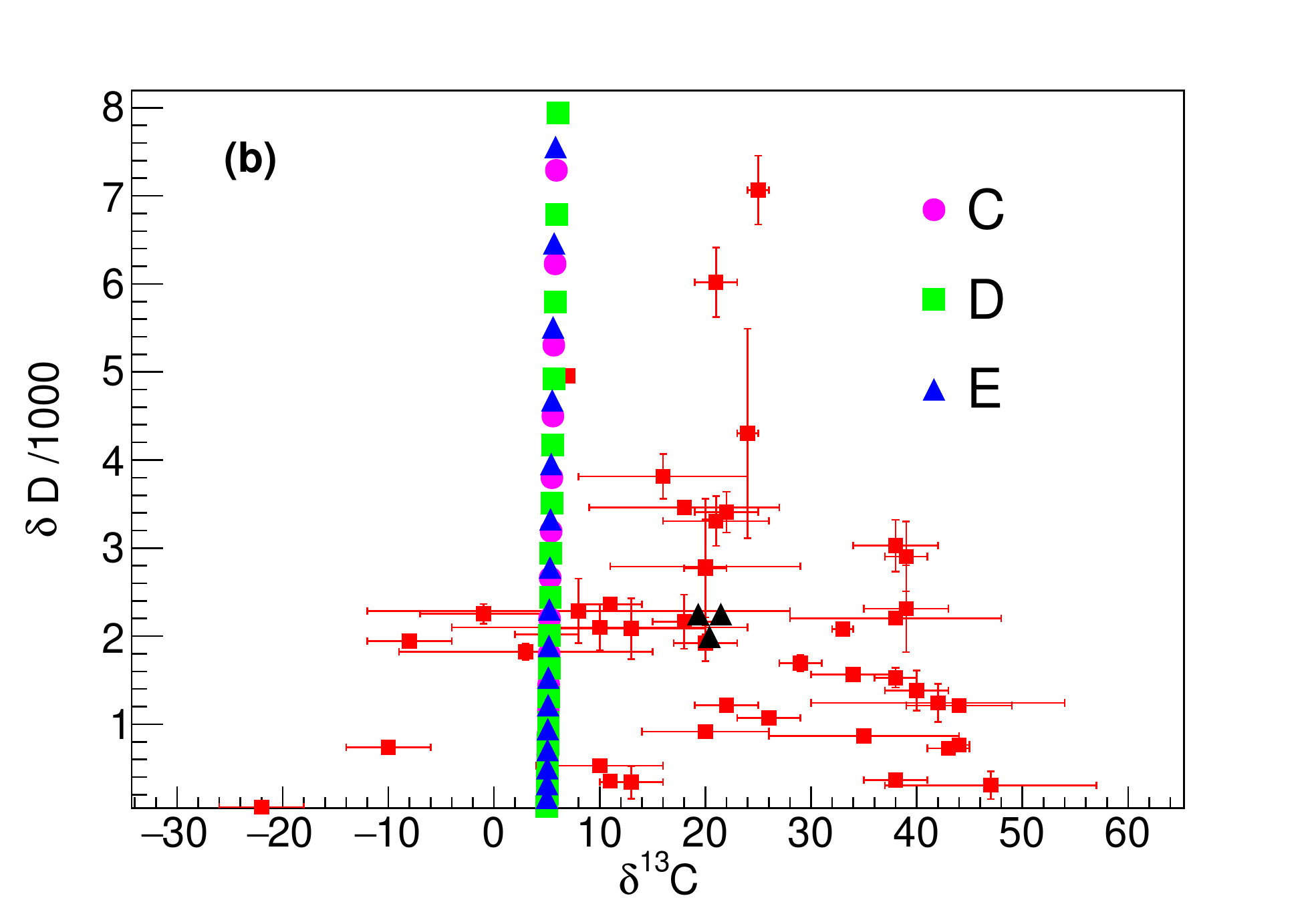}
      \includegraphics[width=0.49\textwidth]{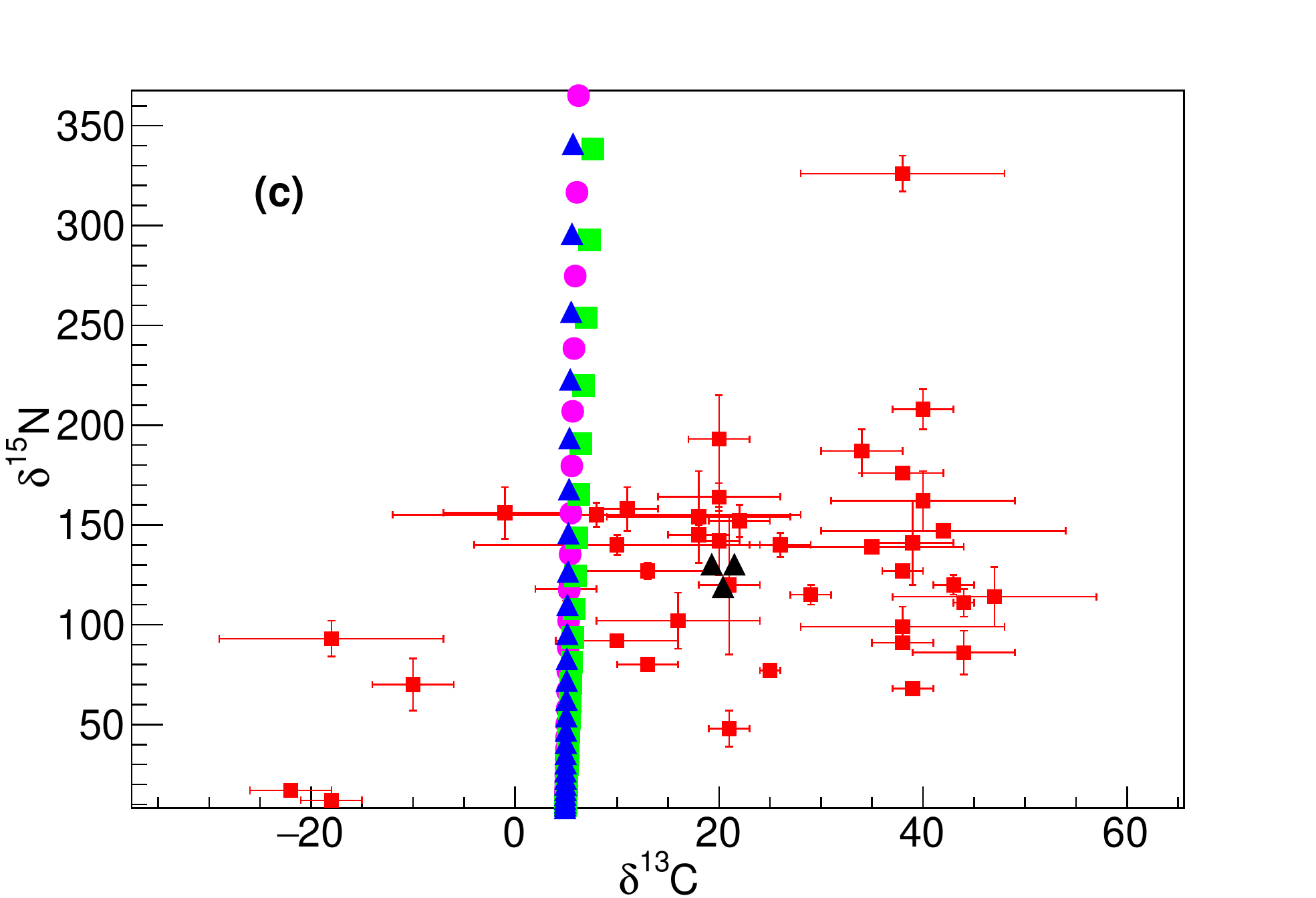}
    \caption{Isotopic abundance ratios
    assuming an initial abundance equal to
    those of the Orgueil meteorite~\cite{famiano19}.  Shown are (a) $\delta$D vs. $\delta^{15}$N, (b) $\delta$D vs. $\delta^{13}$C, and (c) $\delta^{15}$N vs. $\delta^{13}$C.  The red squares represent individual meteoric measurements, while the other symbols falling along lines represent calculational results.  The black trefoil in each panel is the average value of the measurements.}  Each symbol of the calculation corresponds to a different integrated neutrino flux, with integrated flux increasing from the lower left to the upper right.  Used with permission
    \cite{famiano19}.
    \label{isotope_ratios}
\end{figure}

While the results are presented in a prior work~\cite{famiano19}, an example is shown in  Figure \ref{isotope_ratios} compares the calculated isotopic 
anomalies with data of Elsila et al.~\cite{elsila12}. Qualitatively, the 
representation of the data is quite reasonable, given that the $\delta D$
values average around several thousand, those for $\delta^{15}N$ around
150, and those for $\delta^{13}C$ around 20 at about 0.1 to 0.5 seconds (although this apparently 
depends on whether the analyzed sample was taken from an inclusion or from
the body of the meteorite, as discussed further below). As noted in Elsila et al.~\cite{elsila12}, the
extremely high $\delta D$ values are for specific amino acids: the $\alpha$-amino acids.

Presumably the amino acids that would be most likely to form from the atomic and molecular soup
would dominate the abundances of the amino acids following recombination. However, this would
undoubtedly depend on the different atoms and molecular fragments that were produced not just from
the anti-neutrino interactions on $^{14}$N but on all the atoms in all the molecules that existed
prior to the arrival of the anti-neutrinos.

In the absence of information that would lend guidance to the above considerations, 
though, the best we can hope for is qualitative agreement between the calculations and the data.
As discussed above, this does appear to be the case.
Since the amino acids that were not destroyed by the anti-neutrino burst would favor left-handedness, they would guide the subsequent recombination and the autocatalysis discussed below.

As noted above, the D, $^{15}$N, and $^{13}$C abundances are well represented over the two 
orders of magnitude over which they are observed, as seen from the trefoils, representing the 
averages of all the isotopic anomalies observed for the amino acids, in Figure 
\ref{isotope_ratios}. Indeed, two of the Rosswog and Liebend\"orfer scenarios are very close to 
the trefoil indicated average for roughly ten processing intervals. This agreement is especially 
interesting because it depends on details of the calculations that might not be apparent at the 
outset. Most notable, perhaps, is the sensitivity of the results to the iron group nuclides, which
have been found to exist at small abundance levels in the meteorites, but which have huge neutron 
capture cross sections. Not including these would have resulted in a $\delta$D that was much too
large for all three scenarios of Rosswog and Liebend\"orfer \cite{rosswog03}, and which surely
would miss the averages of the amino acid isotopic anomalies by a large amount.

Note the scales of the axes in Figure \ref{isotope_ratios}.
The qualitative result of this study is that nuclear processes are able to change D abundances
by several orders of magnitude, $^{15}$N abundances by $\sim$0.2, and $^{13}$C by $\sim$0.01 as reflected in the data.
Nuclear processes are seen to produce large variations in the isotopic ratios for D, $^{15}$N, and very little variation in the $^{13}$C ratio.  We note
carefully that the isotopic ratios in $^{15}$N extend up to about 35\%, while that of D extends up to several orders of magnitude, and variations in $^{13}$C
are only on the order of a few percent, where the scatter in the points exceeds the average value.  The computations in isotopic ratios cover similar ranges.
It can be seen, however, that the calculated variation in the $^{13}$C ratios is very small compared to the overall scatter in the data.  
This is discussed in a previous
paper~\cite{famiano19} in which model and environmental uncertainties may account for uncertainties in the computations.  
For example,
the calculations presented here do not include effects from aqueous alteration processes~\cite{sephton13,glavin09}.   Further, this model seems to reproduce data more closely
matching the measurements of more primitive meteorites~\cite{famiano19}.  Subsequent processing may further alter $^{13}$C/$^{12}$C ratios in
meteoritic environments.
\section{Autocatalysis Triggered by Weak Interactions in Nuclei}
\label{autocat_sec}
Current models of the formation of amino acid chirality rely heavily on 
some mechanism of autocatalysis in order to increase the enantiomeric excess 
of biomolecules in a mix.  In many cases, the autocatalysis is treated 
as a subsequent mechanism which takes place after a non-zero enantiomeric
excess is formed.  

Here, two autocatalysis models \cite{Frank1953,farshid17} are explored in
terms of coupling the SNAAP model to subsequent or concurrent autocatalysis.  In
one of the models explored, the SNAAP model processing occurs prior to the auto-catalysis.
However, it will be shown that autocatalysis may occur subsequent to the
processing mechanism for one of these models.  For the other model, autocatalysis
must occur simultaneous to some mechanism which continually drives
the mixture towards homochirality (though that mechanism need not be
the initial mechanism creating a non-zero enantiomeric excess, and it may
even be stochastic~\cite{farshid17}).

While the models explored here use selective destruction from the SNAAP
model as a trigger or driver for subsequent auto-catalysis, these models can
be adapted to any situation in which an enantiomeric excess is induced by any
mechanism.  In fact, in one of the models explored here, the originally assumed 
mechanism is stochastic in nature~\cite{farshid17}.  We replace the
stochastic mechanism with a deterministic driving mechanism.

While the original SNAAP model assumed a very high magnetic field and a very
high anti-neutrino flux, we have generalized the model to evaluate the effects of autocatalysis in
a much lower field with lower anti-neutrino flux.  This serves the purpose of
relaxing many of the original constraints of the SNAAP model, increasing the number possible scenarios in
which it may occur, and - in many cases - reducing the possible violent nature of sites
originally thought to induce chirality.  In what follows, we take a simplistic approach
in assuming that the SNAAP mechanism and autocatalysis operate concurrently.  While this is necessary
in one of the autocatalysis models, autocatalysis may occur
subsequent to chiral selection in the other.   This is pointed out later.
\subsection{Autocatalysis With Antagonism}
Because meteoritic $ee$s are sometimes larger than those predicted for the SNAAP model except in extreme circumstances, or for any other model that has been suggested, we examine a possible autocatalysis mechanism which includes selective amino acid destruction via weak interactions in magnetic
fields (nuclear processes) combined with chemical processes of amino acid creation and destruction.  The 
process for creation and destruction of spin-states in amino acids is dependent on chirality as shown 
previously \citep{famiano18}.  This is combined with a modified version of the Frank model \citep{frank53}, which relates these processes to the molecular chirality.  
In combination, all spin/chiral states 
are coupled.

The reactions for the SNAAP model were described above.  The autocatalysis model described in this section,
which is derived from the original Frank model~\cite{frank53,ja19},
is mediated by the following \textit{chemical} reactions:
\begin{subequations}
	\label{reactions}
\begin{eqnarray}
A + D &\xrightarrow[]{k_a} 2D\\
A + L &\xrightarrow[]{k_a} 2L\\
D + L &\xrightarrow[]{k_n} 2A\\
X &\xrightarrow[]{k_c} A
\end{eqnarray}
\end{subequations}
where the rates are indicated for each interaction above the arrow.  The labels D and L indicate 
right-handed and left-handed molecules respectively, and A indicates an arbitrary molecule interacting with 
each chiral molecule.  The last reaction above is used to indicate the overall creation of species A at a rate
$k_c$, where X is any process that creates A.

The above molecular reactions were coupled to the nuclear reactions of the SNAAP model.  In this analysis, each
species is a combination of spin state and chiral state, where the number abundances are indicated by
$D_{\phi}$ and $L_{\phi}$ for left-handed and right-handed chiral states respectively,
and $\phi =\pm$ indicates the spin state relative to the external magnetic field \cite{famiano18}.

We note that the reaction sequence of Equations \ref{reactions} represents a mathematical model, and 
an extant reaction sequence replicating these reactions has eluded chemists for nearly four decades~\cite{blackmond20}.
While the source term has been added as a modification to the above sequence, the true puzzle comes from
the self-replicating nature of reactions \ref{reactions}a and \ref{reactions}b.  Such a sequence could have a 
basis in the famous Soai reaction~\cite{soai90,soai92,soai94,soai95,soai95b}.  Self-catalyzing processes
in Equation \ref{reactions}a and \ref{reactions}b have also been shown to proceed in the proteingogenic amino acids 
via crystallization~\cite{viedma08}.
(For a good review of the Frank model, we refer to citations in \cite{blackmond20,ja19,saito13}.)  It is also important
to note that reactions \ref{reactions}a and \ref{reactions}b represent a specific reaction or reaction sequence.  While not a
reactions maintain the chirality of the original reactants, a specific reaction is necessary to do this.

We also note that in the original Frank model, the growth in abundance of one particular enantiomer does not necessarily have 
a well-defined
steady-state~\cite{ja19}.  In the original model, Equation \ref{reactions}c was $D+L\rightarrow\o$, in which the destruction
of chirality by D and L enantiomers proceeded via an arbitrary pathway.  This has been addressed in prior work~\cite{ja19} by
limiting the total concentration of the catalyst $A$ with the reaction in Equation \ref{reactions}c.  However, we do
add a source term to this sequence which could account for the slow creation of a catalyst via an arbitrarily reaction represented by
Equation \ref{reactions}d.   We explore the sensitivity of this sequence to this source term below.

Because molecular chiral states and nuclear spin states are coupled, Equations \ref{reactions} must be expanded to include 
spin states as well.  
For all reactions, the initial spin states of the particles can be specified.    
However, as the species A is
arbitrary in this model, its spin state cannot be specified; here, it is assumed to be spinless.
To ascribe a spin state to A, one would then simply create an additional species, $A\rightarrow A_\pm$.  
If the spin state of A 
is the nuclear spin state as assumed in the SNAAP model, A would necessarily have
a nuclear spin.  In this case, however, it is assumed that the spin states of the products are in
thermal equilibrium.  We will discuss the ramifications of this choice in the next section.

In the SNAAP model nuclear spin states are coupled to the molecular chirality through the nuclear magnetic shielding 
tensor.  However, the mathematics developed here can be applied to any model in which a two-state system within the molecule (besides
chirality) is coupled to the molecular chiraltiy.  Using the example of the SNAAP model, in which different chiral states are
destroyed at different rates, the
net rate equations for chiral destruction, coupled to autocatalysis and chiral antagonism are then:
\begin{widetext}
\begin{subequations}
	\label{diff_eq}
\begin{equation}
\dot L_+ = (\varepsilon - R_p)L_{+} + \delta L_{-} - k_nL_+D + 2k_af_+AL-k_aAL_{+}
\end{equation}
\begin{equation}
\dot L_- = -\varepsilon L_{+} - (\delta + R_a) L_{-} - k_nL_-D + 2k_af_-AL-k_aAL_{-}
\end{equation}
\begin{equation}
\dot D_+ = (\varepsilon - R_a)D_{+} + \delta D_{-} - k_nLD_+ + 2k_af_+AD-k_aAD_{+}
\end{equation}
\begin{equation}
\dot D_- = -\varepsilon D_{+} - (\delta + R_p) D_{-} - k_nLD_- + 2k_af_-AD-k_aAD_{-}
\end{equation}
\begin{equation}
\dot A = -k_aA(L+D) + 2k_nLD + k_c
\end{equation}
\end{subequations}
\end{widetext}
where the first two terms on the right side of Equations \ref{diff_eq}a-d are the the quantities linking spin states to chirality
which mediate destruction of one spin state over another via the destruction rates $R_p$ and $R_a$.  
In terms of the SNAAP model, the destruction rates
are from weak interactions with the nitrogen nuclei \citep{famiano18}.
Number densities without spin-state subscripts correspond to
total number density by taking the sum of both spin states (e.g., $L = L_+ + L_-$).  In the above rate equations, $R_p$ is the 
rate of destruction of nuclei spin-aligned with the local magnetic field, and $R_a$ is the rate of destruction of nuclei
spin-anti-aligned with the local magnetic field.  
In the SNAAP model explored previously, these values are very close to
each other, and typical values of $1-R_a/R_p\sim 10^{-5}$ -- $10^{-6}$ have been found
based on the averaged shielding tensor values \citep{famiano18,famiano18b}.  The rates $k_a$, $k_d$, and $k_n$ are the rates for the reactions indicated
in Equation \ref{reactions}.  The terms, $\epsilon$ and $\delta$ are derived from
the thermal spin equilibration rates of molecules:
\begin{eqnarray}
\varepsilon &= \frac{\Delta f-1}{2T_1}
\\\nonumber
\delta & = \frac{\Delta f + 1}{2T_1}
\\\nonumber
\Delta f &= f_+ - f_-
\end{eqnarray}
where $T_1$ is the longitudinal nuclear spin relaxation time in the the external field and the terms $f_\pm$ are the fractions of nuclei
with magnetic moment $\mu$ in spin-aligned and spin-anti-aligned states in a thermal population:
\begin{eqnarray}
f_+&\equiv \frac{\exp{\frac{\mu B}{kT}}}{\exp{\frac{\mu B}{kT}}+1}
\\\nonumber
f_-&\equiv \frac{1}{\exp{\frac{\mu B}{kT}}+1}
\end{eqnarray}
Here, we treat the nuclei as only having two spin states, though the model is applicable to nuclei with more than two states.

\begin{figure}
    \centering
     \includegraphics[width = 0.48\textwidth]{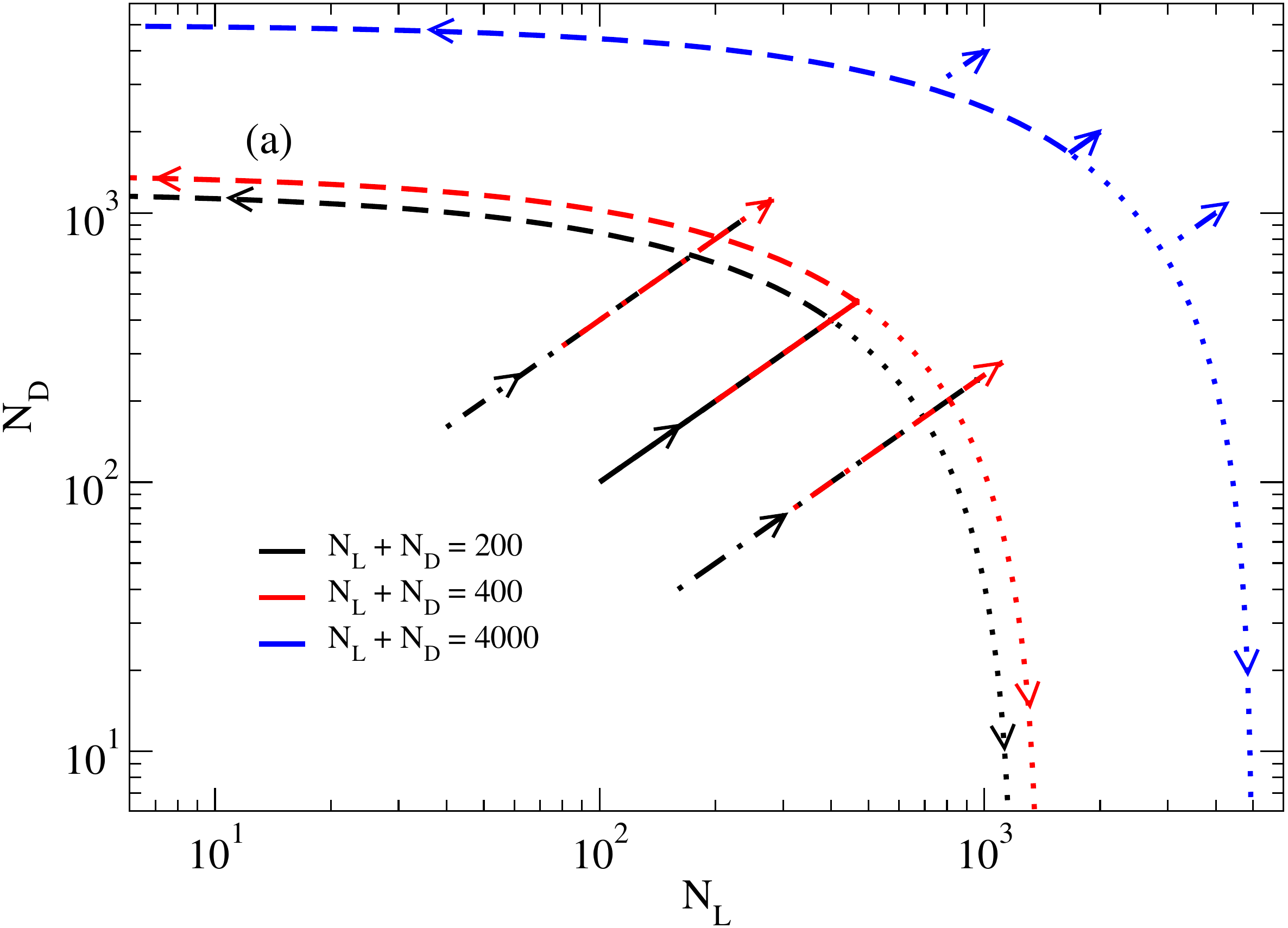}
     \includegraphics[width = 0.49\textwidth]{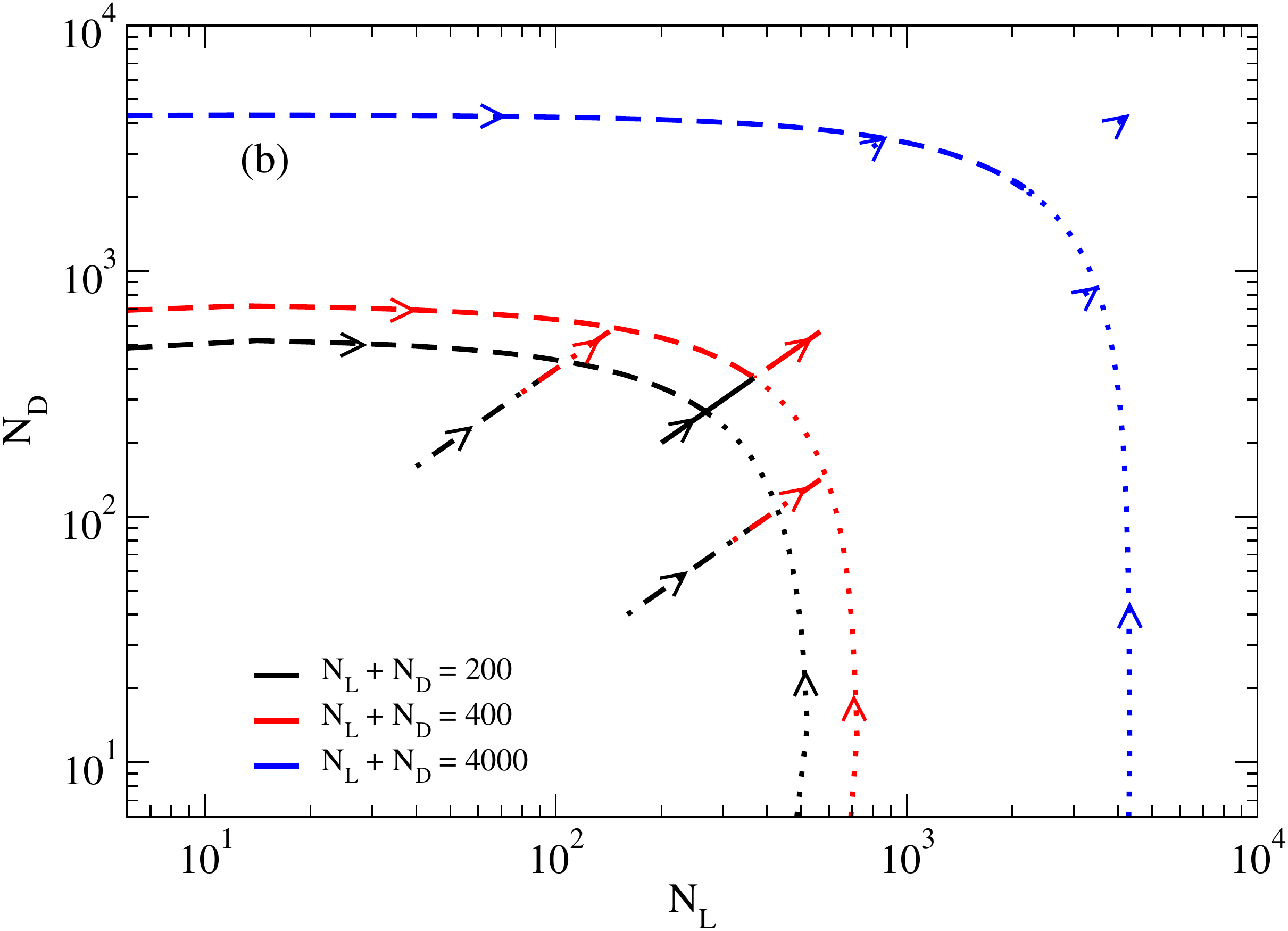}
    \caption{
        Stability of each autocatalysis model explored in this work. The color of each line indicates the initial abundances specified in the legend with $A(t=0)=1000$.  The solid lines correspond to conditions with 
        the initial $ee=0$ ($N_L = N_D$) for each.  The dot-dash lines in each case correspond to an initial $ee=\pm0.67$ for various initial populations and $k_n=0$.  The arrows indicate the time evolution in each case.
        (a) Phase diagram of the autocatalysis + antagonism model for various initial conditions.  The dotted line in each case corresponds to an initial very small positive $ee$.  The dashed lines correspond to an initial very small negative $ee$.
        (b) Phase diagram for the autocatalysis only model.  The dotted lines in each case correspond to an initial positive $ee$ very close to 1, while the dashed lines correspond to an initial negative $ee$ very close to -1.}
    \label{phase_dia}
\end{figure}
The stability of this model is shown in Figure \ref{phase_dia} (Left).  
Here, the evolution of enantiomeric abundances, $N_L$ and $N_D$, is
shown.  In this figure, the selective destruction rates are zero, $R_p = R_a = 0$, and any evolution is a result of the autocatalysis process only.  The abundance evolution for a specific set of initial abundances was then followed.  For this particular figure, the rates in 
Equations \ref{diff_eq} are set to $k_a=k_n=k_c=10^{-7}$.  The initial abundance of the species $A$ is set at $A=1000$.  The arrows indicate the direction of the abundance evolution in time.

Several scenarios are shown in this figure. These include cases with initial
enantiomeric excesses of zero and cases with non-zero initial enantiomeric excesses.   For the cases with an initial enantiomeric excess of zero, indicated by the solid lines, it can be seen that, though the abundance changes for L and D enantiomers due to interactions with
species A, the enantiomeric excess does not change, $N_L = N_D$ always.

However, if the system starts off with an intial non-zero enantiomeric excess, the system evolves towards a homochiral mixture.  This is shown by the dotted lines for a slightly positive enantiomeric excess, and the dashed lines for a slightly negative initial enantiomeric excess.
The dotted lines correspond to initial abundances $(N_L,N_D)$ of
(400.00001, 399.99999), (500.00001, 499.99999), and (2000.00001, 1999.99999), resulting in very small initial $ee$s.  The values are
reversed for the slightly negative initial $ee$s, with 
the evolution shown by the dashed lines.

The dot-dash lines in Figure \ref{phase_dia} correspond to
initial non-zero $ee$s, but with a non-autocatalytic
rate $k_n=0$.  In this case, the calculations all start with
an initial $ee=\pm0.67$ for the initial abundances indicated in the legend.  As expected, with $k_n=0$, there is no antagonism, and the production rates of D and L enantiomers proceed at the same rate, increasing slightly owing to the production provided by the presence of A.

\begin{figure}
    \centering
     \includegraphics[width = 0.49\textwidth]{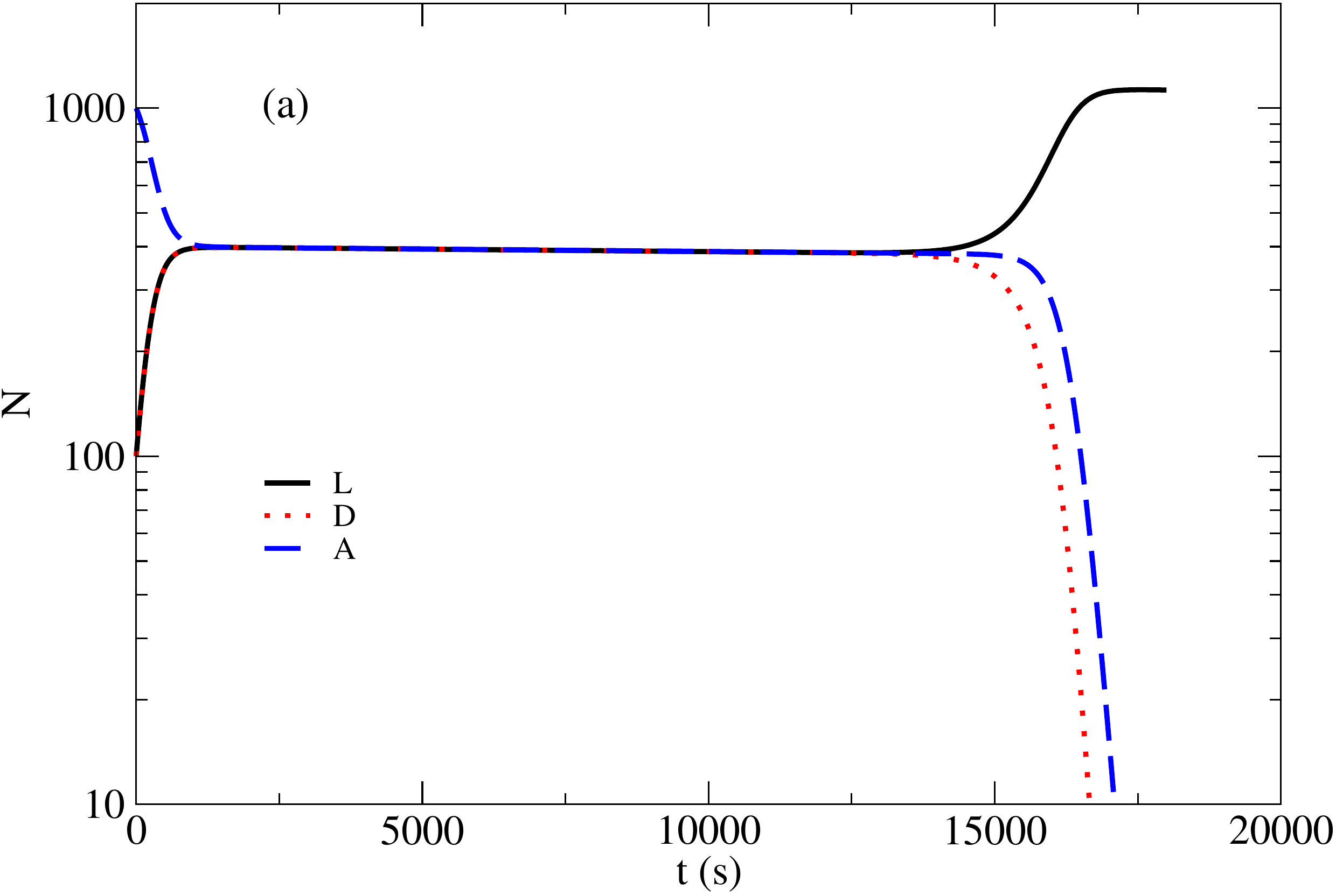}
     \includegraphics[width = 0.49\textwidth]{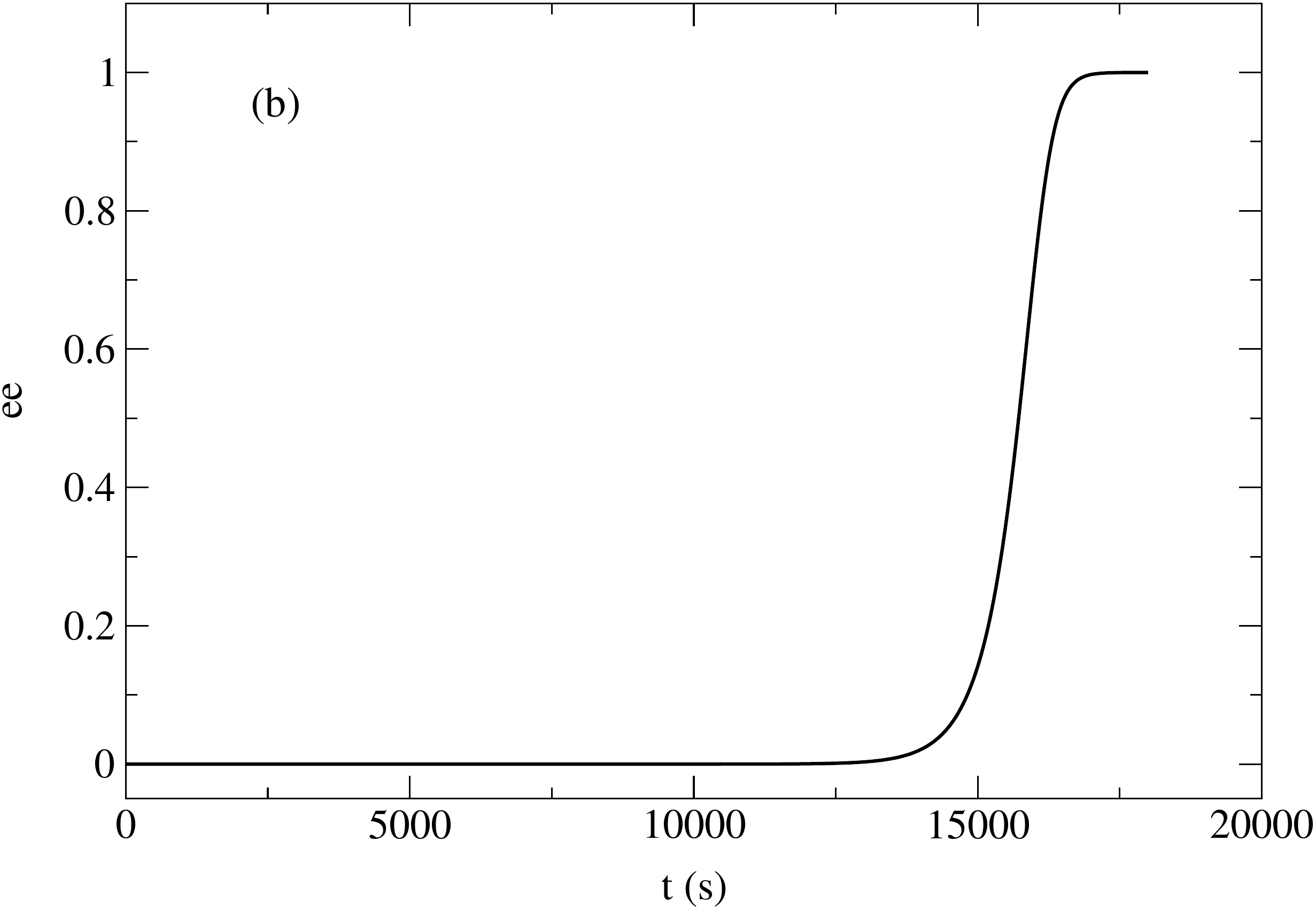}
    \caption{
        Results for an antagonistic autocatalytic model corresponding to the SNAAP model with $k_a=k_c=k_d = 10^{-7}/(2T_1)$.  This is an adaptation of
        the SNAAP model with $R_a = 10^{-7}/(2T_1)$ and $(R_a - R_p)/R_a = 10^{-6}$ in an external field of 1 T.
        Shown are (a) the number of each species as a function of time and (b) the enantiomeric excess as a function of time.     The solid, dotted, and dashed lines
        correspond to the abundances $L$, $D$, and $A$ respectively.
        }
    \label{ee_gr}
\end{figure}

Results from a simple evaluation are shown in Figure \ref{ee_gr}.  Here the $ee$ as a function of time is shown for a low field, B = 1 T, and low 
weak interaction rate as indicated in the figure caption.  Here, we
set $L(t=0)=D(t=0)=100$ in thermal spin populations with $A(t=0)=1000$. As seen in the figure, given enough time $ee$s can become extremely large, approaching 100 percent. The assumed parameters are not especially extreme; large $ee$s can be obtained over a wide range of parameters.

\begin{figure*}
    \centering
    \includegraphics[width=0.49\textwidth]{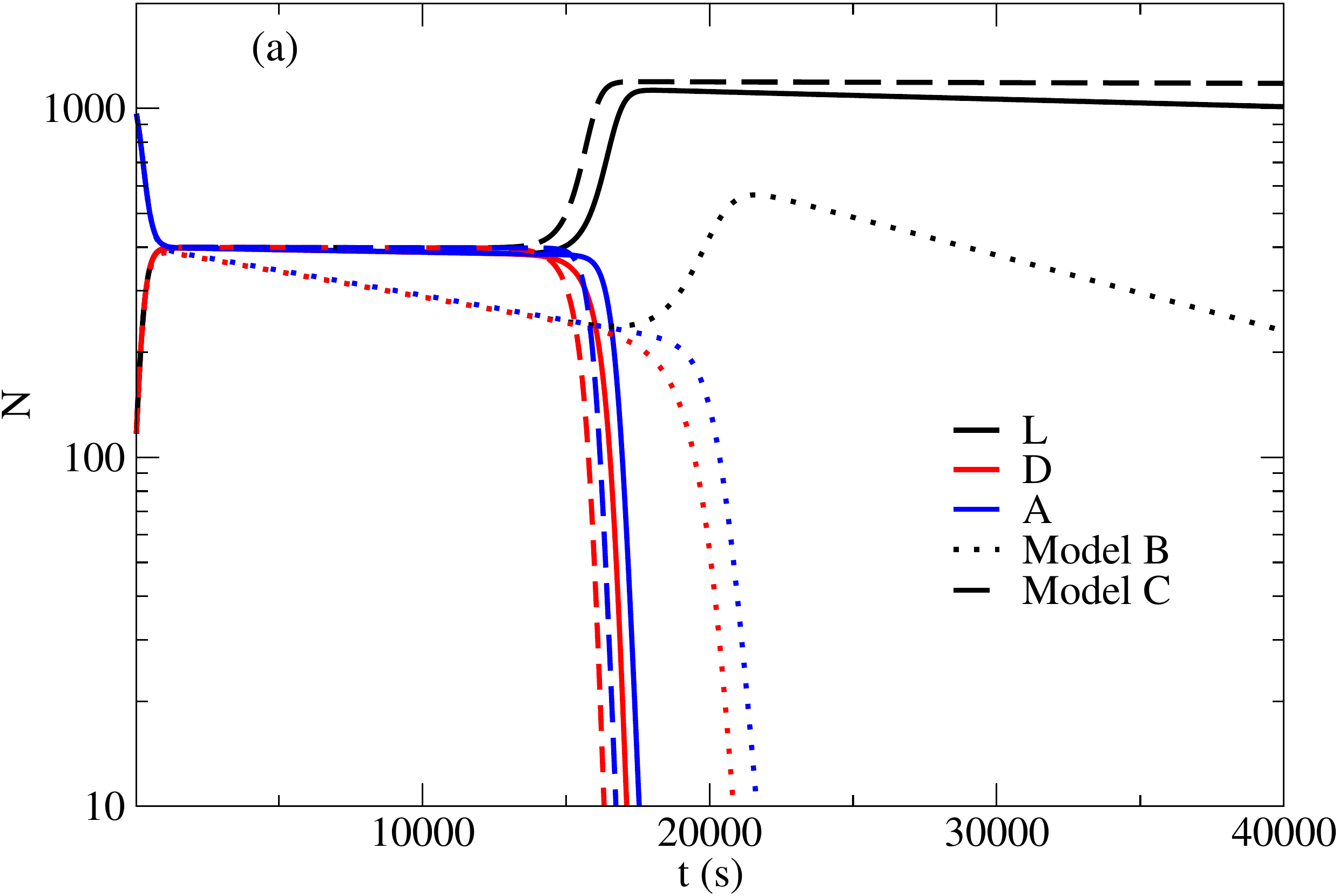}
    \includegraphics[width=0.49\textwidth]{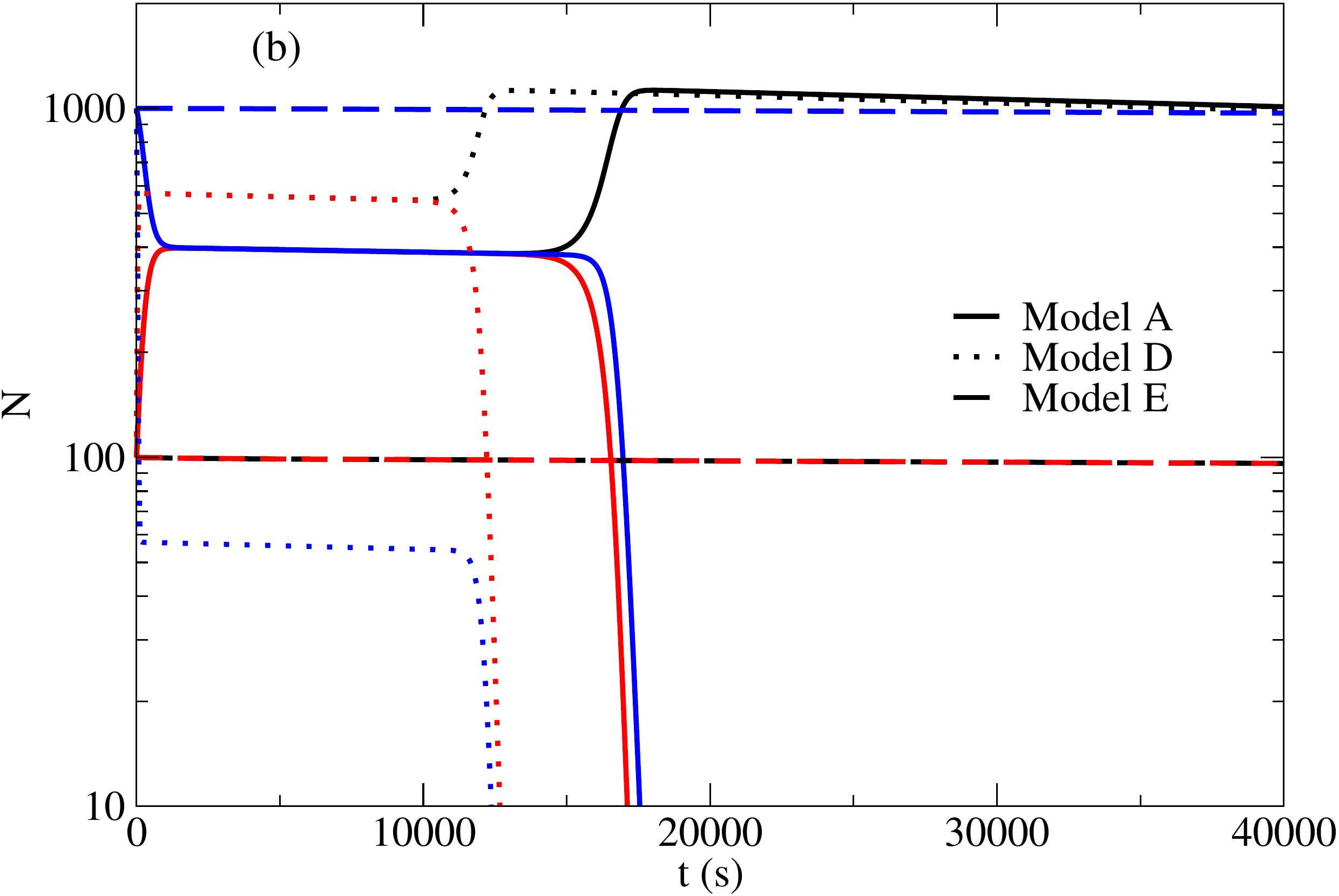}\\
    \includegraphics[width=0.49\textwidth]{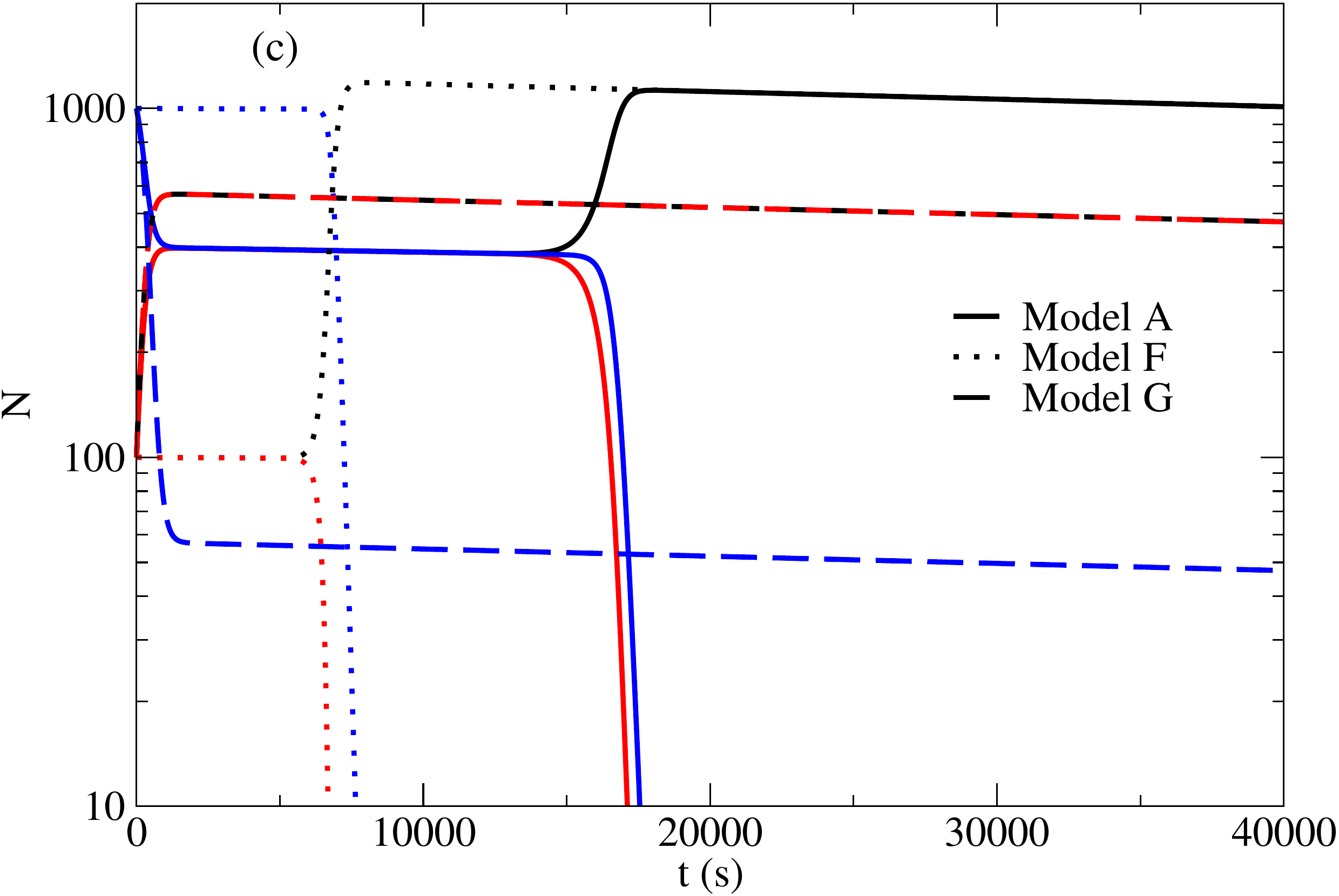}
    \includegraphics[width=0.49\textwidth]{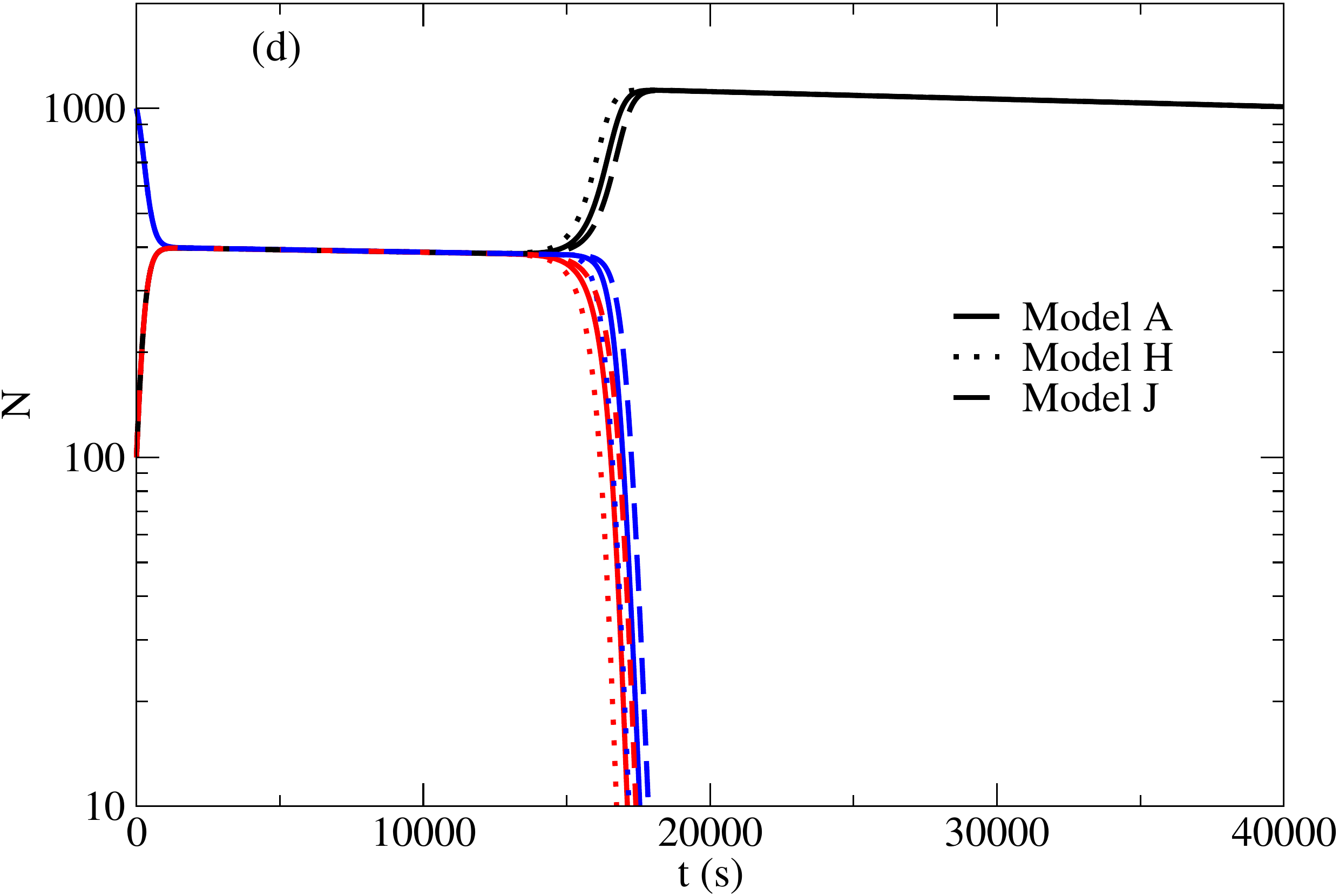}\\
    \includegraphics[width=0.49\textwidth]{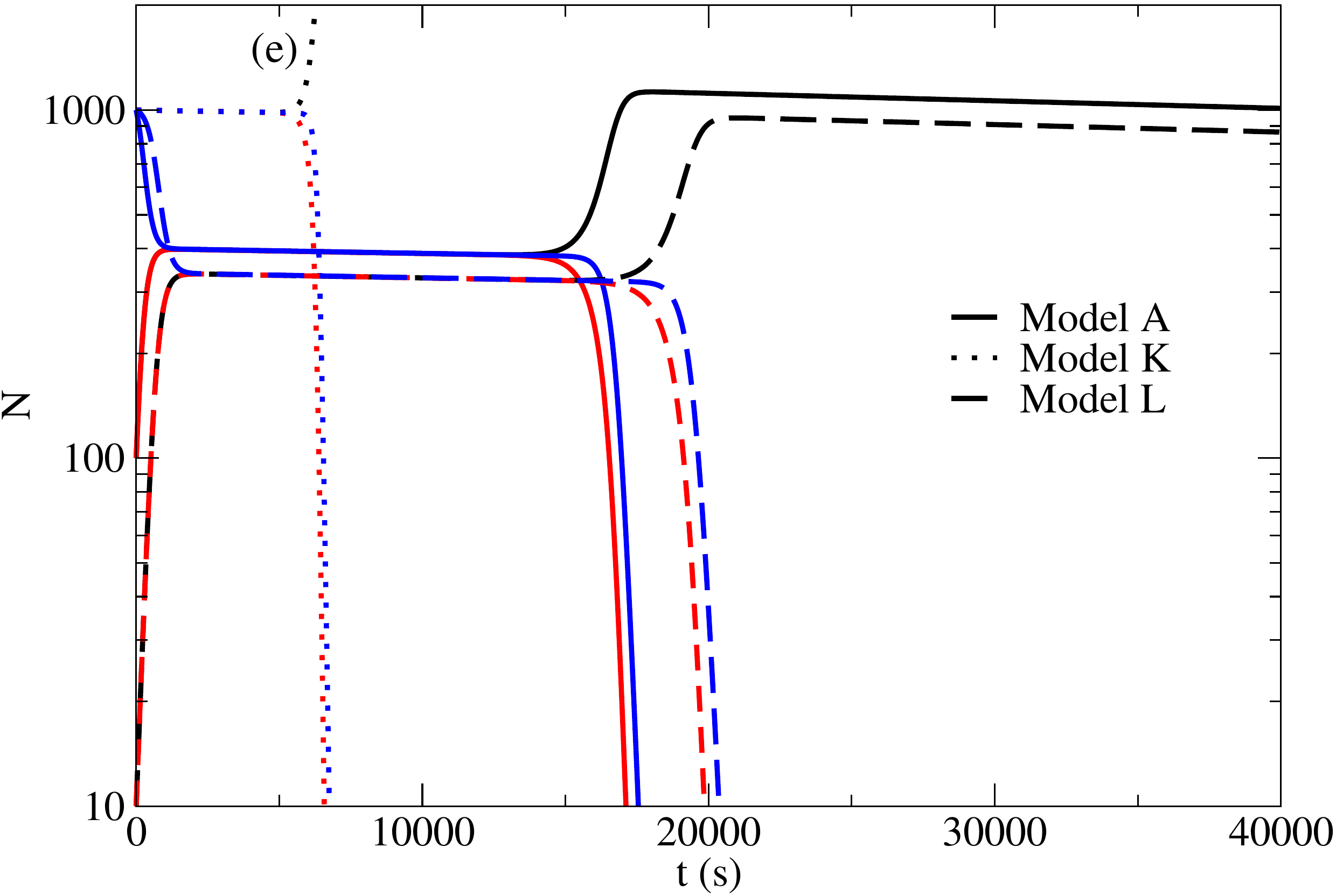}
    \includegraphics[width=0.49\textwidth]{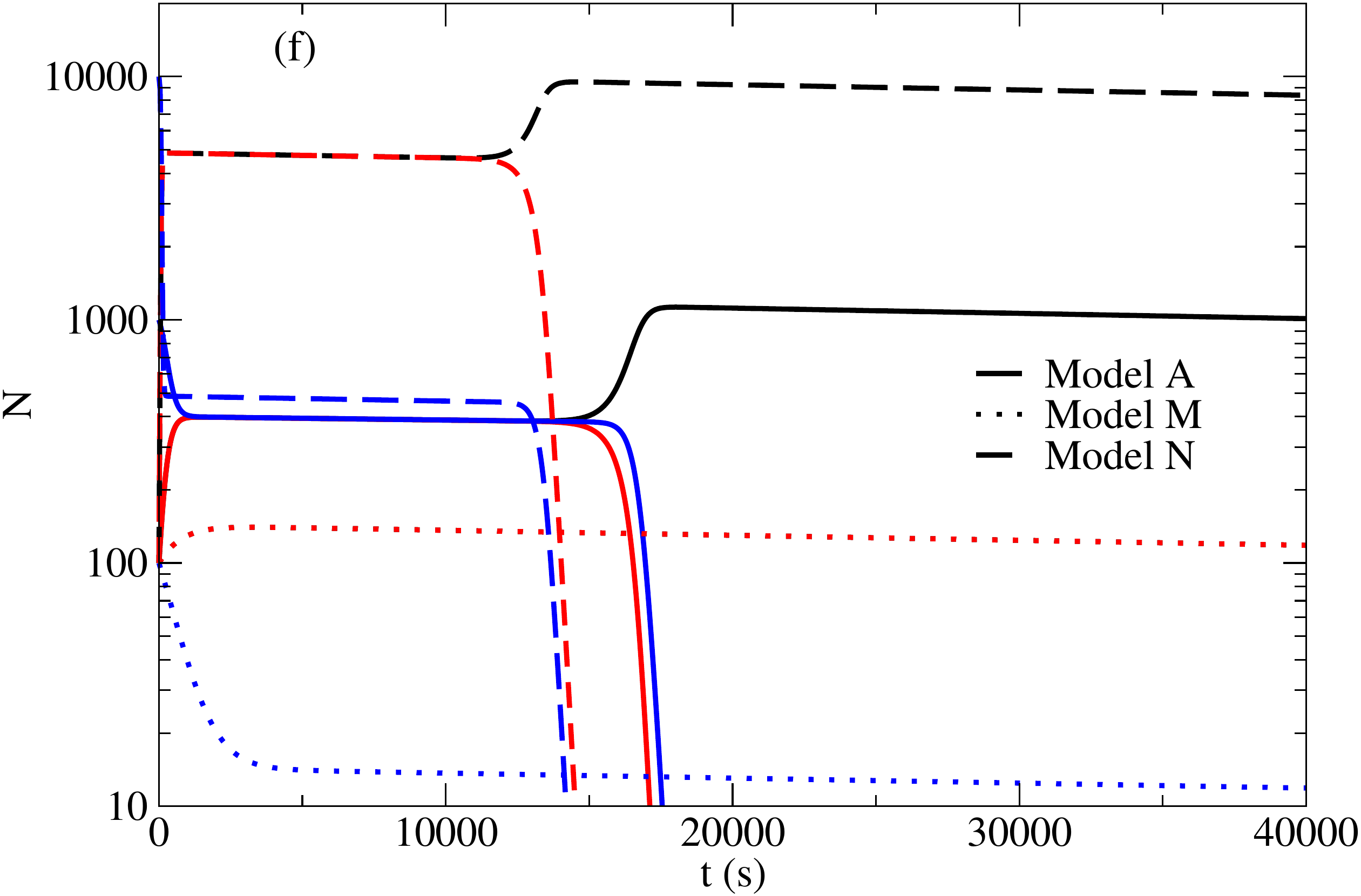}
    \caption{Comparison of different model results for various parameters
    of the autocatalysis-plus-antagonism model.  The parameters
    associated with each model are indicated in Table \ref{frank_params}.  The black lines correspond to L-enantiomers; the red line corresponds to D-enantiomers, and the blue line corresponds
    to the additional catalyst $A$ indicated in equation \ref{diff_eq}e.  For each panel the solid line
    corresponds to Model A, and the dotted and dashed lines are indicated in each figure.}
    \label{frank_mod_comp}
\end{figure*}

Several models were chosen to predict $ee$s using the autocatalysis-plus-antagonism scenario. These are shown in Table \ref{frank_params}.  The starting model, Model A, is indicated in this table, and
only parameters which differ from those of model A are shown in the table for
subsequent models.  The time evolution of each enantiomer's abundance is shown in Figure
\ref{frank_mod_comp}.
The number is each isomer is shown for L- and D- enantiomers as well as the additional species A
as black, red, and blue lines respectively.  The model type is indicated by solid lines for model A and dotted and dashed lines for the other models.  

Two major characteristics which are explored in this figure are the time to which a maximum
enantiomeric excess is created and the absolute values of the abundance
of each enantiomer.  The time to which a maximum (or minimum) enantiomeric abundance is achieved is found to depend heavily on the overall
autocatalysis rate, the non-autocatalytic rate, and the initial 
number abundance of the amino acids.  It depends to a lesser extent on the 
overall destruction rate and the creation rate of the catalyst.

The magnitudes of the abundances of individual enantiomers is found to depend 
more heavily on the non-autocatalytic rate and the initial number abundances of all
molecules, but to a much lower extent on the destruction rate, the autocatalytic
rate, and the catalytic creation rate.
\begin{table}
    \centering
    \begin{tabular}{|c|c|c|c|c|c|c|c|}
    \hline
    \textbf{Model}&
    \textbf{$R_a\times2T_1$}&\textbf{$k_a\times2T_1$}&\textbf{$k_n\times 2T_1$}&\textbf{$k_c\times 2T_1$}&
    \textbf{$N_0$}&\textbf{$A_0$}\\
    \hline
    A& $10^{-7}$ &  $10^{-7}$ & $10^{-7}$ & $10^{-7}$  & $10^2$ & $10^3$\\\hline
    B& $10^{-6}$ & ~ & ~ & ~ & ~ & ~\\\hline
    C& $10^{-8}$ &  ~ & ~ & ~ & ~ & ~\\\hline
    D& ~ &  $10^{-6}$ & ~ & ~ & ~ &  ~\\\hline
    E& ~ &  $10^{-8}$ & ~ & ~ & ~ &  ~\\\hline
    F& ~ &  ~ & $10^{-6}$ & ~ & ~ &  ~\\\hline
    G& ~ &  ~ & $10^{-8}$ & ~ & ~ &  ~\\\hline
    H& ~ &  ~ & ~ & $10^{-6}$ & ~ &  ~\\\hline
    J& ~ &  ~ & ~ & $10^{-8}$ & ~ &  ~\\\hline
    K& ~ &  ~ & ~ & ~ &  $10^3$ & ~\\\hline
    L& ~ &  ~ & ~ & ~ &  $10$ & ~\\\hline
    M& ~ &  ~ & ~ & ~ &  ~ & $10^2$\\\hline
    N& ~ &  ~ & ~ & ~ &  ~ & $10^4$\\\hline
    \end{tabular}
    \caption{Various parameter sets chosen for the model explored in
    the autocatalysis-plus-antagonism model.  Model A is the base
    model with its associated parameters.  For the other models,
    parameters not indicated are those corresponding to Model A.}
    \label{frank_params}
\end{table}

Here, we note that even a small initial $ee$ can be driven to a significant value (even to homochirality) via antagonistic autocatalysis.  
This may not be surprising given limited results from laboratory experiments~\cite{soai95,soai02,soai14}, which have been shown to autocatalyze mixtures with
initial $ee\sim$10$^{-4}$ to near homochirality. While the SNAAP model is able to produce enantiomeric excesses $\sim$10$^{-4}$, it is possible that 
subsequent autocatalysis may be responsible for increasing the enantiomeric excess.

Caution must be taken when attempting to draw a correlation between the istopic ratios mentioned in a prior paper~\cite{famiano19} and 
apparently large $ee$s produced in meteorites. Because meteoritic enantiomeric excesses may be produced by subsequent autocatalytic mechanisms, 
it is possible that this process is decoupled from the mechanisms responsible for non-zero $\delta$X values.  Thus, we note that - while the results of this 
calculation are compelling, multiple unrelated processes may be responsible for the overall description of amino acids in meteoric environments.  It is
certainly possible that the model described here is a ``trigger'' for subsequent autocatalysis while providing a mechanism for isotopic abundance shifts, but it may
not be totally responsible for producing the bulk enantiomeric excesses seen in meteorites.
\subsection{Autocatalysis Without Antagonism}
We examine an alternate scheme in which the racemic mixture is a global minimum in the rate equations, and subsequent 
deviations from this state will push autocatalysis \citep{farshid17}.  In this model, originally developed to take stochastic
deviations into account, the procession reactions are:
\begin{subequations}
	\label{react_auto}
	\begin{eqnarray}
	A + D &\xrightarrow[]{k_a} 2D\\
	A + L &\xrightarrow[]{k_a} 2L\\
	A &\xrightleftharpoons[k_d]{k_n} L\\
	A &\xrightleftharpoons[k_d]{k_n} D\\
	X &\xrightarrow[]{k_c} A
	\end{eqnarray}
\end{subequations}
This is a modification of the auto-catalysis model of \cite{farshid17} where reaction \ref{react_auto}e was added to introduce constant creation
of species $A$.

The corresponding rate equations for the autocatalysis-only system are:
\begin{widetext}
\begin{subequations}
	\label{auto_eqs}
	\begin{eqnarray}
		\dot L_+& =& (\varepsilon - R_p)L_{+} + \delta L_{-} + 2k_af_+AL-k_aAL_{+} - k_dL_+ +k_nf_+A
		\\
		\dot L_- &=& -\varepsilon L_{+} - (\delta + R_a) L_{-} + 2k_af_-AL-k_aAL_{-}- k_dL_- +k_nf_-A
		\\
		\dot D_+ &=& (\varepsilon - R_a)D_{+} + \delta D_{-} + 2k_af_+AD-k_aAD_{+} - k_dD_+ +k_nf_+A
		\\
		\dot D_- &=& -\varepsilon D_{+} - (\delta + R_p) D_{-} + 2k_af_-AD-k_aAD_{-} - k_dD_- +k_nf_-A
		\\
		\dot A &=& -k_aA(L+D) - 2k_nA + k_d(L+D) + k_c
	\end{eqnarray}
\end{subequations}
\end{widetext}

In the case of the autocatalysis-plus-antagonism model, the racemic state
(L = D) is a point of unstable equilibrium, and a deviation from this
point drives
the system towards an enantiopure state. For the autocatalysis-only
system, if $k_n=0$, the system is maintained in equilibrium at constant
$ee$; any selective destruction 
mechanism can change the $ee$, and a continuous mechanism will drive the
$ee$ towards a new equilibrium state.  If $k_n >0$, then the racemic
state is at equilibrium, and a 
continuous selection mechanism is necessary to overcome the decay of
species $A$. This can be seen by the decay reaction in Equation
\ref{react_auto}c,d in which the decay
of species $A$ into L and D states at equal rates necessarily produces a
racemic mixture.  For an enantiomeric excess to be produced, the
enantiomeric decay rate must exceed 
the decay of $A$ \citep{farshid17}.

The stability of the autocatalysis-only model is shown in Figure
\ref{phase_dia} (right).  For this representation, rates are
$k_a = 3\times 10^{-5}$, $k_n = 2\times 10^{-5}$, $k_c = 0$,
and $k_d = 0.02$. In this figure, the lines and colors 
have the same meaning as on the left side of the figure. However, in the 
case of the non-zero initial $ee$ cases (dashed and dotted lines), the 
initial $ee$s are nearly $\pm$1.  For nearly left-handed homochiral
mixtures, initial values of $(N_L,N_D)$ are (500, 0.0001), (700,0.0001), 
and (4000,0.0001) with the initial values reversed for a nearly 
right-handed homochiral mixture. 

It is seen from this figure that the autocatalysis-only model is stable
for a racemic mixture, and the antagonistic model is unstable for a
racemic mixture.  As expected, in both cases, the system is stable at any $ee$ if 
$k_n = 0$.

\begin{figure}
	\centering
		\includegraphics[width=0.47\textwidth]{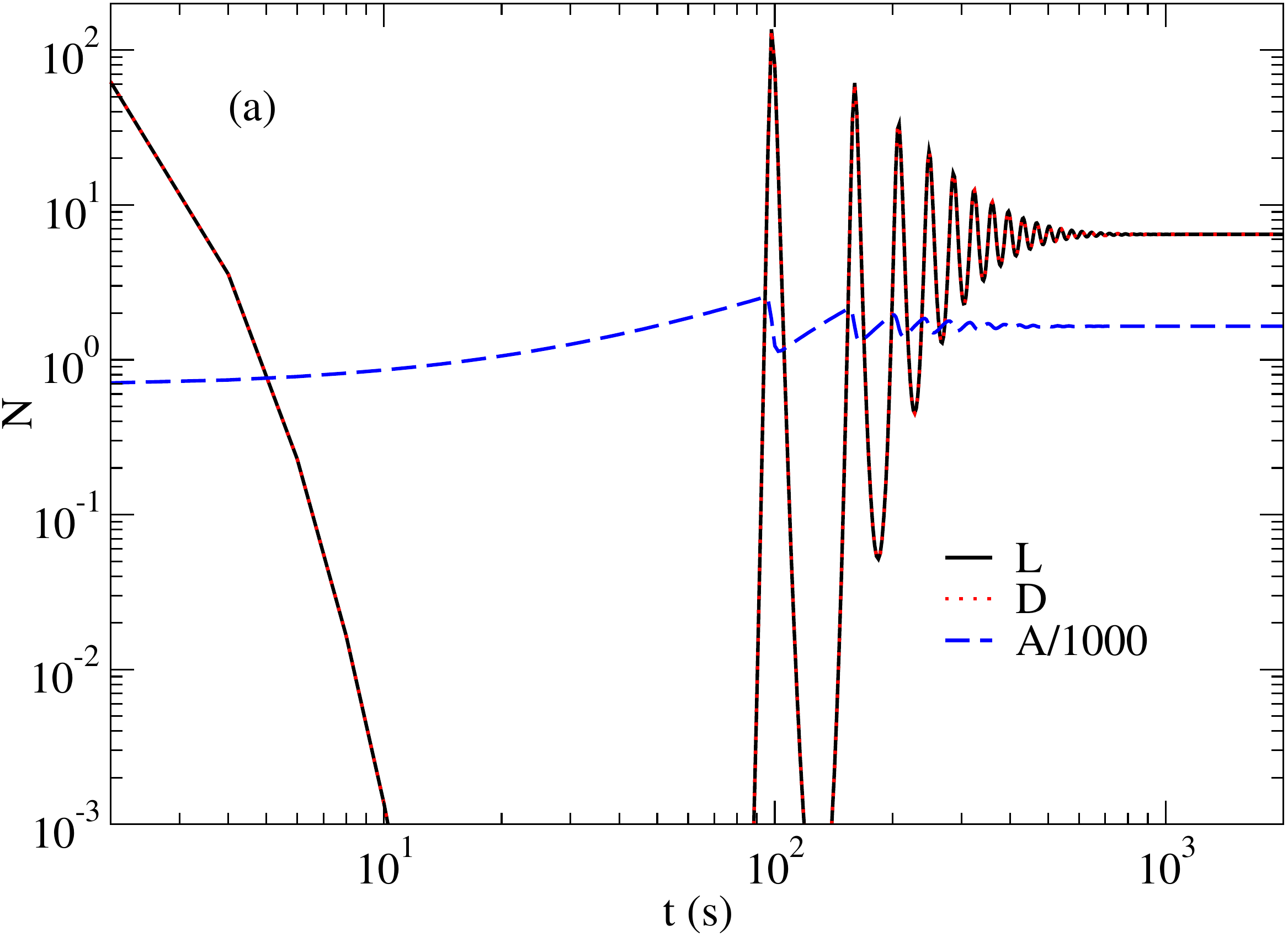}
		\includegraphics[width=0.47\textwidth]{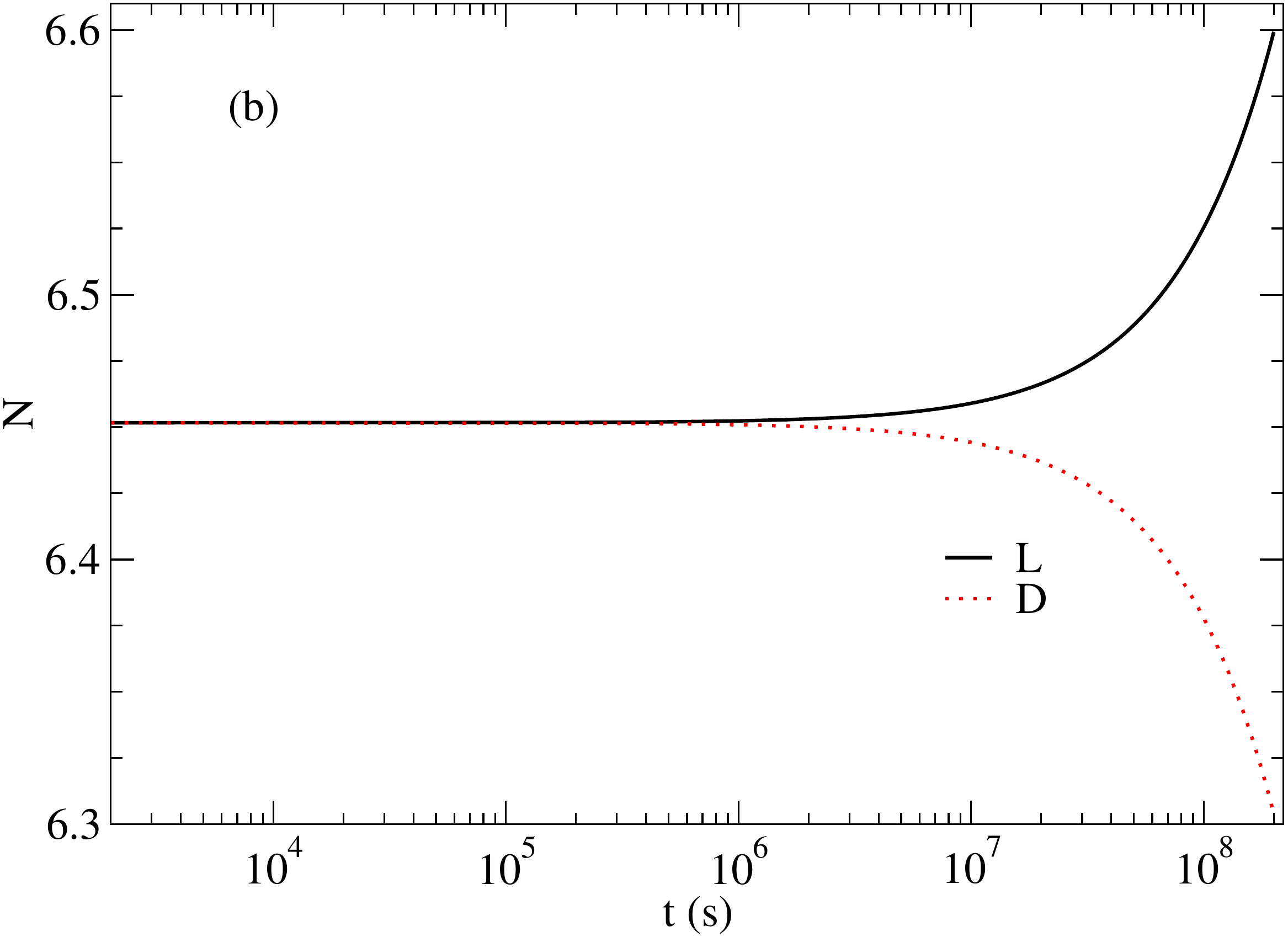}
	\includegraphics[width=0.49\textwidth]{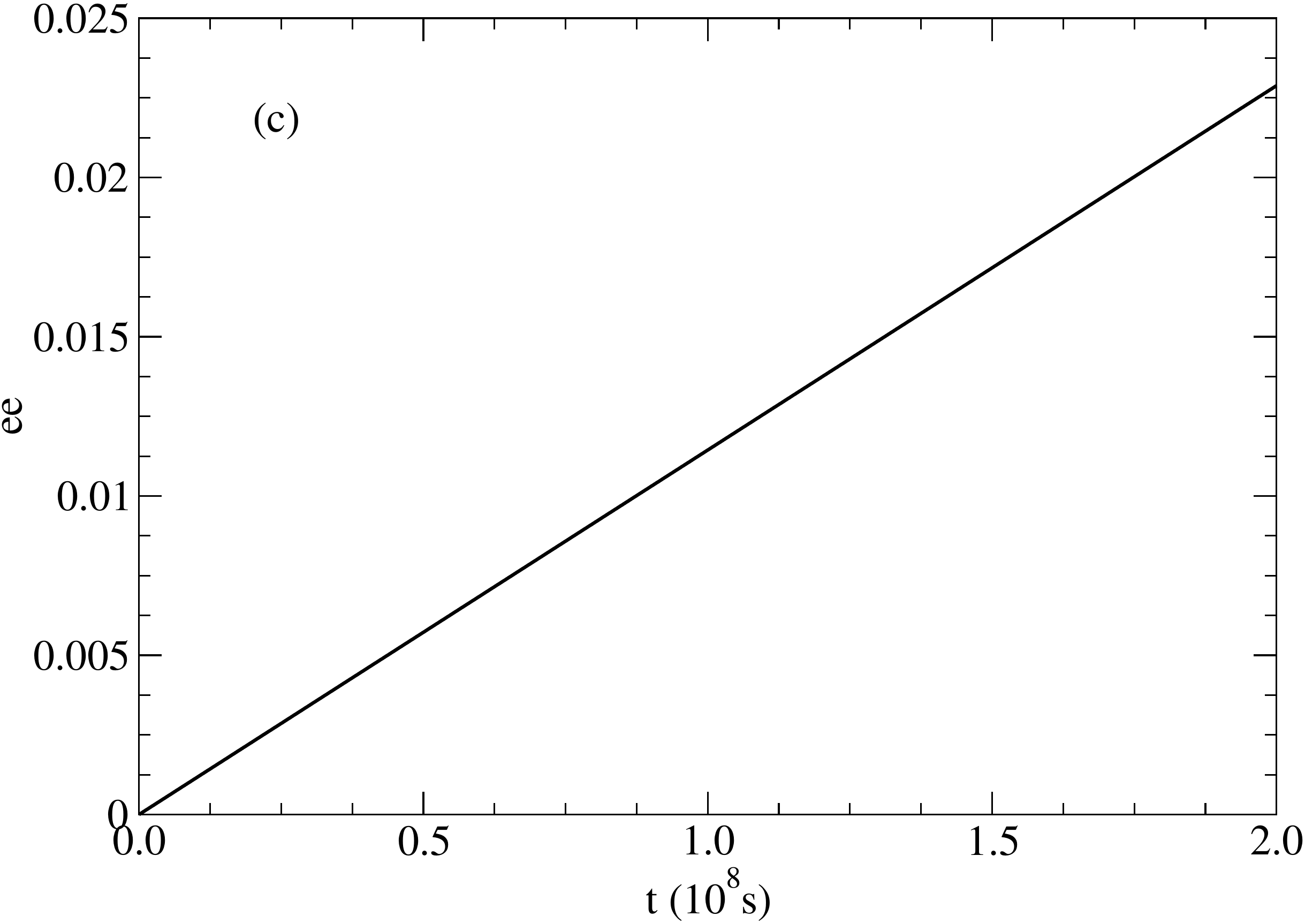}
	\caption{
	Results for an auto-catalysis only model.  In this model, the magnetic field is 10 T, f=0.031$/(2T_1)$, $k_a$=3$\times 10^{-5}/(2T_1)$, $k_n = 0$, $k_c = 0.4/(2T_1)$, and $k_d=0.02/(2T_1)$. Plot (a) shows populations of L, D, and A at low times to indicate initial equilibration, while plot (b) shows populations at high times.  Plot (c) shows the $ee$ as a function of time.    
	}
	\label{ee_auto}
\end{figure}

An example of a simulation showing populations and enantiomeric excess, $ee$, as a function of time for a low-field, medium interaction rate scenario is shown in Figure 
\ref{ee_auto}.  For this calculation, Equations \ref{auto_eqs} are solved using an implicit Bader-Dueflhard method for t$<$1800 s and an explicit Runge-Kutta Prince-Dormand 
method for t$>$ 1800 s.  Initial abundances are $L=D=A=1000$ at t=0.   

Initially, abundances oscillate about equilibrium points as $A$ is destroyed via reactions \ref{react_auto}a and b.  (Note that in this initial model $k_n$=0.) Conversely, $A$ is
created via reactions \ref{react_auto}c-e.  After equilibrium, it is seen that the $ee$ grows linearly due to the selective destruction from the SNAAP mechanism.  If the system 
evolves long enough, the enantiomeric excess is expected to approach unity assuming that the destruction mechanism continues.

In a manner similar to the autocatalysis-plus-antagonism model, variations in the modified autocatalysis-only model were explored.
Various models studied are shown in Table \ref{jafa_params}.  Different parameter sets were chosen to show how these affect the production as a function of time in this model.  (Note that for the autocatalysis-only models, model labels are
lower case, while in the autocatalysis-antagonism model, labels are in upper case).
\begin{table}
    \centering
    \begin{tabular}{|c|c|c|c|c|c|c|c|}
    \hline
    \textbf{Model}&
    \textbf{$R_a/(2T_1)$}&\textbf{$k_a/(2T_1)$}&\textbf{$k_d/(2T_1)$}&\textbf{$k_c/(2T_1)$}&
    \textbf{$N_0$}&\textbf{$A_0$}\\
    \hline
    a& 0.01 &  $10^{-5}$ & 0.01 & 0.5  & $10^3$ & $10^3$\\\hline
    b& 0.001 & ~ & ~ & ~ & ~ & ~\\\hline
    c& 0.1 &  ~ & ~ & ~ & ~ & ~\\\hline
    d& ~ &  $10^{-6}$ & ~ & ~ & ~ &  ~\\\hline
    e& ~ &  $10^{-4}$ & ~ & ~ & ~ &  ~\\\hline
    f& ~ &  ~ & 0.001 & ~ & ~ &  ~\\\hline
    g& ~ &  ~ & 0.1 & ~ & ~ &  ~\\\hline
    h& ~ &  ~ & ~ & 0.05 & ~ &  ~\\\hline
    j& ~ &  ~ & ~ & 5.0 & ~ &  ~\\\hline
    k& ~ &  ~ & ~ & ~ &  $10^2$ & ~\\\hline
    l& ~ &  ~ & ~ & ~ &  $10^4$ & ~\\\hline
    m& ~ &  ~ & ~ & ~ &  ~ & $10^2$\\\hline
    n& ~ &  ~ & ~ & ~ &  ~ & $10^4$\\\hline
    \end{tabular}
    \caption{Various parameter sets chosen for the model explored in
    the autocatalysis-only model.  Note that for the autocatalysis-only
    models, the model parameters are lower-case, while in
    the autocatalysis-plus-antagonism model, the model labels are in upper case.  Model a is the base
    model with its associated parameters.  For the other models,
    parameters not indicated are those corresponding to Model a.}
    \label{jafa_params}
\end{table}

\begin{figure}
    \centering
    \includegraphics[width=0.49\textwidth]{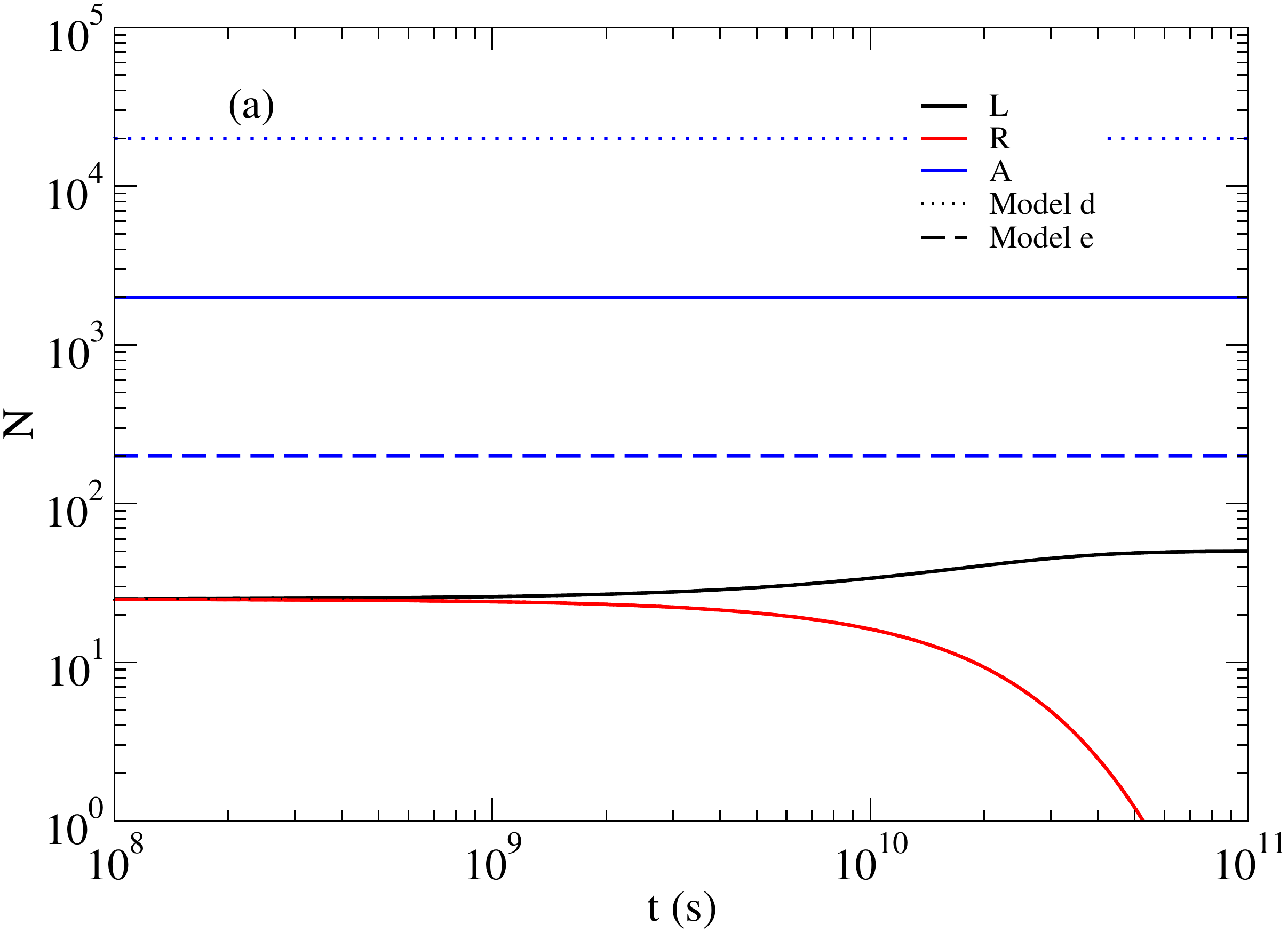}\\
    \includegraphics[width=0.49\textwidth]{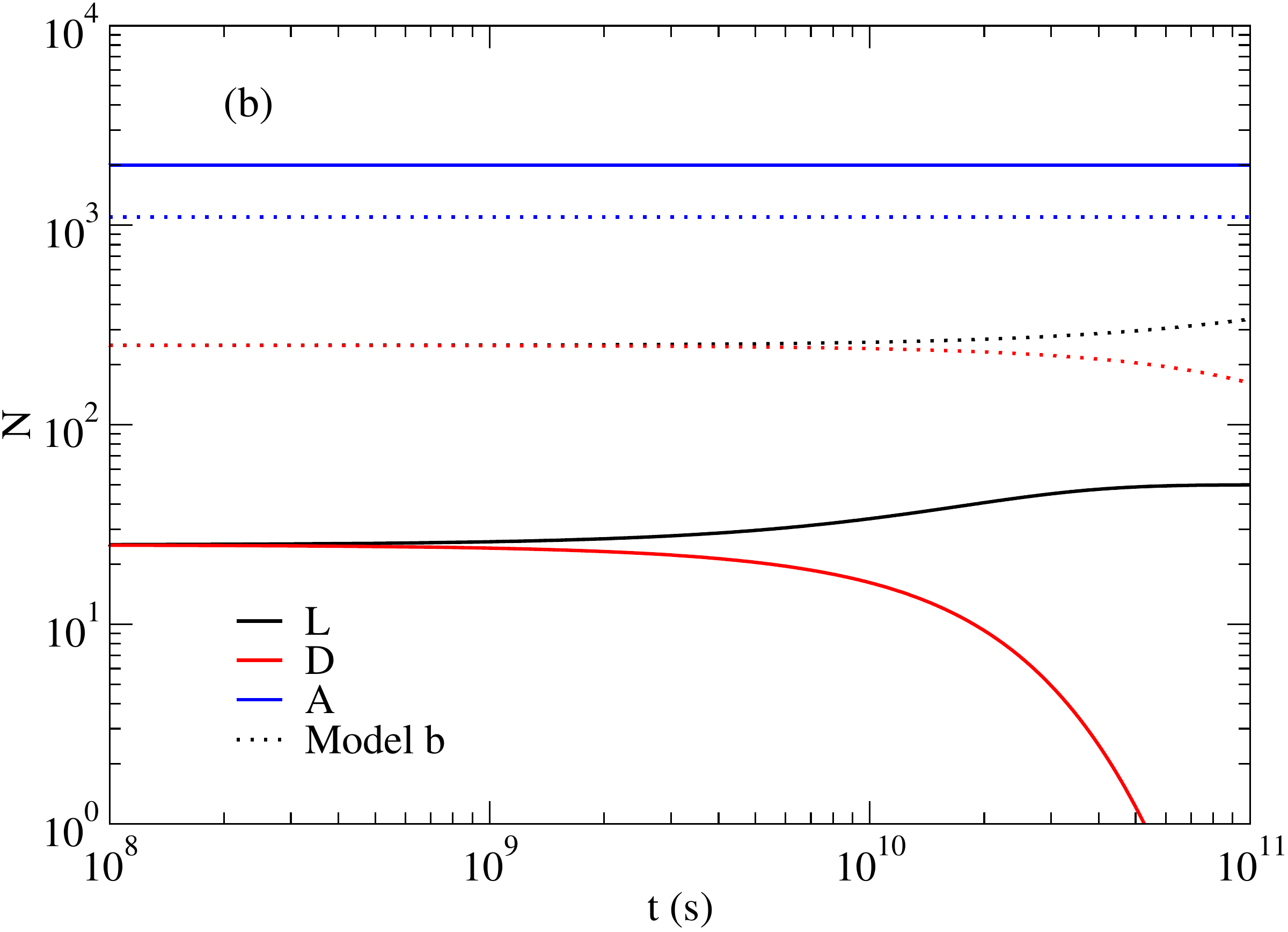}
        \includegraphics[width=0.49\textwidth]{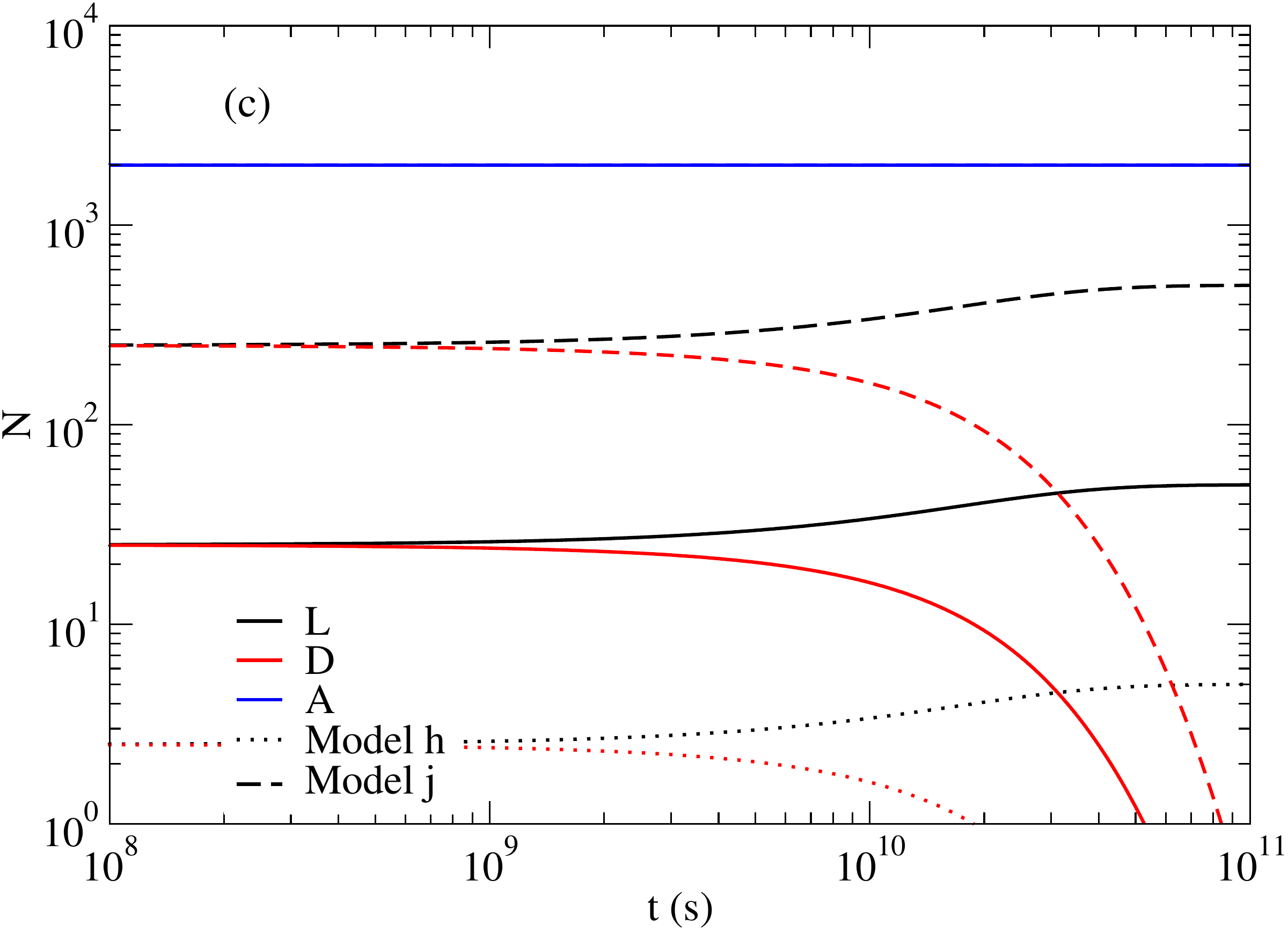}

    \caption{Model comparison for models a, b, d,e, h, and j. The line styles and colors are the same as those in Figure \ref{frank_mod_comp}.}
    \label{ao_models}
\end{figure}

Some results of these various models are compared in Figure \ref{ao_models}.  
For Model b, the lower overall destruction rate results in a higher amino acid
abundance.  However, since it is the destruction rates $R_a$ and $R_p$ which push
the enantiomeric excess away from zero, the time to a homochiral mixture is longer.  Model c, on the other hand, is not visible in this figure as the destruction rate
is much more rapid, and the amino acids are not replaced quickly enough to form an
appreciable abundance.  In this case, L- and D- amino acids are destroyed quickly (within a time scale not visible on the figure.)

A similar effect was noted in model d, in which the autocatalysis rate is too low to maintain production of individual enantiomers, resulting in
in an overall desctruction of the amino acids with a corresponding growth of species A.

In models h and j, the overall time to homochirality is the same in all cases
with a different final abundances of each enantiomer owing to the differences
in creation rates of A.  Finally, it is found that changing the initial abundances
of each enantiomer has a small effect on the time to which a significant $ee$ is 
formed, but has little effect on the overall net $ee$.

For models k -- n, the initial abundances are altered.  However, equilibrium creation and destruction rates are approached rapidly, and the long-term behavior of the model remains unchanged.

It was also found that the low amino acid decay rates used in this 
treatment, relative to the creation rate, resulted in little effect
on the overall evolution of the abundances.

\subsection{Spin-State Considerations of The Reactants and Products}

In this model, it was assumed that the daughter products in the auto-catalysis reactions
are produced in equilibrium spin states. In other words, the thermodynamic equilibration
time of the
spins of the $^{14}$N nuclei is much shorter than
the chemical reaction time.  This assumption is likely safe as the rates
chosen in the models evaluated here are assumed to be much lower than the 
spin relaxation time for $^{14}$N.  If the reaction
rates are assumed to be much greater than the spin relaxation times, then assumptions
must be made about the spin states prior to and after the reactions.  The thermal 
equilibrium production of spins is reflected in the rate equations by multiplying
the total enantiomeric abundances in the products by the equilibrium fractions $f_{+,-}$.

It is also emphasized that the autocatalysis model defined here assumes two
coupled two-state systems.  One two-state system is the molecular
chirality.  The other two-state system is subject to the specific model
incorporated into the auto-catalysis mechanism.  In the case of the SNAAP
model, the nuclear spin is coupled to the molecular chirality.  In other models,
a variety of nuclear and molecular properties may be incorporated, including
molecular magnetic moment (or induced magnetic moment), electric dipole moment, charge state, or virtually any property of the molecule.  Of course, not all combinations result in a truly chiral effect \cite{barron00}. 

This is noted here as the SNAAP model relies on the coupling of nuclear spin
to the molecular chirality via an external magnetic field \cite{famiano18,famiano18b,famiano19,boyd18,boyd18a}. In the idealized case, the
spin of interest is that of the $^{14}$N in the amine group.  However,
the ground-state $^{14}$N nucleus has a spin of 1, giving it three possible spin
states in a magnetic field.  So far, only the spin-aligned and spin-anti-aligned 
states are treated, while the spin-perpendicular state is ignored.  Since the nuclear destruction in this model
is based on the spin-alignment, the spin aligned, and spin-anti-aligned states represent extremes in the spin alignment relative to the neutrino momentum
vector in the SNAAP model.  The spin-perpendicular state is effectively averaged
into the spin-aligned and anti-aligned states, simplifying the model somewhat, though including this third state is possible and would be a relatively simple
extension of the general model presented here. It is ignored for simplicity and its inclusion is not expected to make a significant difference in the overall results as its destruction rate would simply be an average of the spin-aligned and spin-anti-aligned cases.  

\section{Conclusions}
The results of comparisons of theoretical studies of $ee$s and isotopic anomalies in meteorites within the SNAAP model are subject to significant uncertainties, in both the data and the calculations.  
	Many of these have been explored in prior work, along with the sensitivity of the resultant enantiomeric excesses to these uncertainties~\cite{famiano18,famiano18b,boyd18}.
	Typical uncertainties in measurements of the $ee$s can be as large as 50\% ($\sigma$)~\cite{burton18}, while the data can have significant scatter.  For example,
	a compilation of isovaline $ee$ in meteorites shows a large range of values, 0$<$ee$<$20.5\%.  
	In terms of isotopic abundance ratios, the scatter and uncertainties shown in Figure \ref{isotope_ratios} is considerable.  These may indicate 
	significant underlying differences in meteoric environmental and chemical evolution and morphology, some of which has been explored~\cite{elsila12,glavin10}.

Much of the computation involved in the present work has room to mature.  While the autocatalysis model
presented here is an adaptation of prior models, the parameters involved are dependent on the specifics of the
chemistry involved.  The uncertainties of the autocatalysis parameters have been discussed in \S\ref{autocat_sec}, where
sensitivity to the input parameters is explored.
In computing $ee$s and isotopic ratios, perhaps the largest uncertainties are with the microscopic parameters.
While the neutrino interaction cross sections are a significant uncertainty, these predominantly affect the rates at which isotopic abundance ratios change and the
overall rates at which the $ee$ changes in a model.  Thus, a longer exposure can result in similar overall abundance ratios or $ee$s for a lower flux~\cite{famiano18,boyd18}.  
Likewise, the autocatalysis mechanism can enhance even a small $ee$ given an initial imbalance.

Perhaps the most significant environmental parameter is the external magnetic field.  This parameter was found to have a significant effect on the overall 
final $ee$ in the absence of autocatalysis~\cite{boyd18,famiano18}.  However, as was found in the present work, relatively small magnetic fields can 
result in significant $ee$s.  Since autocatalysis mechanisms may operate concurrently or subsequent to the amino acid processing, this expands the current model
in terms of environmental conditions for which amino acids may viably be processed as well as the time over which they may be processed.

We also note that any autocatalysis may be decoupled from both the formation of an initial enantiomeric excess and from the shifts in isotopic abundance
ratios.  While the shifts in isotopic abundance ratios and the initial enantiomeric excess may be caused by the same mechanism, autocatalysis may operate 
independently, and even subsequently to the mechanisms resulting in a shift in $ee$.

 However the qualitative success of the model is unmistakable. It does appear to predict $ee$s comparable to those observed for objects passing close to an extreme cosmic event, most notably, the two NS mergers studied in this work, with plausible parameters. With the autocatalysis that is expected to occur (and is required for all cosmic models of $ee$ production) the $ee$s of a fraction of a percent could achieve the order of ten percent level observed for some amino acids. The SNAAP model also gives a remarkable representation of the isotopic anomalies from which the amino acids and other molecules would be recreated following the anti-neutrino burst and subsequent neutron captures. The model predicts a huge enhancement of D with respect to H, a large enhancement of $^{15}$N with respect to $^{14}$N, and a much smaller enhancement of $^{13}$C with respect to $^{12}$C, all of which are consistent with observations.  

While it is very interesting that the expected reactions together with production of a non-zero $ee$ can
produce isotopic abundances that qualitatively match with those in amino acids
found in meteorites, those isotopic abundances can also be produced
in other ways.  D-enrichment can be related to cold chemistry.
In fact, D-enrichment in comets in our solar system and molecular clouds is understood
in terms of cold chemistry and it must play a role to some extent \cite{cleeves16}.  Also, the formation
process of amino acids, which depends on the type of amino acids, can produce
the molecular species-dependent isotopic abundance pattern that seems to match
with those found in meteorites \cite{elsila12}. 
In the SNAAP model, selective destruction of right-handed molecules, as well as many non-amino-acid molecules, will result in free ions and fragments of those that are destroyed, some of which would be expected to recombine to form new amino acids. Furthermore, any water or ice present in the meteoroid will contribute $^{16}$O, from which $^{15}$N would be made. Since N is the least common of the elements that are essential for forming amino acids, that might be expected to provide a large enhancement of $\delta^{15}$N to enable production of amino acids. Thus, the amino acids most readily formed out of the detritus of the destroyed right-handed molecules and any water would tend to be the ones with the largest isotopic anomalies. They would probably tend to be more racemic than the molecules from which they were produced, unless autocatalysis played a major role. 

It might provide an interesting test of the SNAAP model if the isotopic anomalies were greater in the left-handed molecules than the right-handed ones, although it might be difficult to obtain sufficient statistics to determine this. Perhaps the Hayabusa2~\cite{hayabusa2a,hayabusa2b,hayabusa2c,orex_hayabusa} and 
OSIRIS-REx~\cite{orex_hayabusa,orexa,orexb} missions could provide sufficiently large sample sizes to do so.


The present results also suggest that two NS mergers are a potentially ideal site for producing appreciable 
enantiomeric excesses in nearby meteoroids, although neutron-star-massive-star binaries, where the neutron star has an accretion disk, might also. When the massive star becomes a supernova, it would provide sufficiently energetic antineutrinos to produce 
	the inverse $\beta$ decay of $^{14}$N in the disk and the more remote portions of the disk would not have its preexisting amino acids subjected to the hostile environment produced nearer the supernova.  Likewise, the probability that a binary system with a NS contains a companion that is not
a NS is expected to be much higher.  Recent models suggest that roughly 20\% of current NS binary systems contain a companion that is neither a NS nor a black hole~\cite{tutukov02}.
This latter scenario, if feasible, would be particularly interesting, especially in light recent considerations of habitable planets in orbit about neutron 
stars\cite{patruno17}.  Certainly, more work must be done to expand the current model to non-merger, non-SN events.
Furthermore, both may well explain the enantiomeric 
excesses observed in meteorites, and possibly, even, how the amino acids achieved the left-handedness that is essential for them to contribute to life on Earth. The recent observation of a large $^{60}$Fe abundance in Antarctic ice \cite{ko19} suggests that a major cosmic event did occur close enough to Earth that could have produced both the $ee$s and the isotopic anomalies. 

We note that the autocatalytic times presented in each model here are much longer than the original 
times proposed in the SNAAP model.  However, several factors are relevant to this.  First, the times in 
the SNAAP model are scalable.  That is, while the original SNAAP model relies on a neutrino burst, 
there is nothing to preclude a source of sufficiently high energy neutrinos from a much lower flux source, in which case, the
time to a non-zero enantiomeric excess could be longer.  Also, the models proposed here assume 
autocatalysis that is concurrent with the formation of an enantiomeric excess. It might be possible that the autocatalysis follows the initial enantiomeric excess formation, in which case the timescales
could be dramatically different.  However, we note that the autocatalysis-only model described herein does
require a constant push away from a racemic state to result in a homochiral state.

In the present work, two autocatalysis models are modified to incorporate a 
``trigger'' mechanism which tips the enantiomeric excess towards a non-zero value.  In the autocatalysis-antagonism model, any non-zero $ee$ is driven towards homochirality,
though the mechanism for this is not specified.  In this model, the non-zero $ee$ tends to be produced by the SNAAP model discussed in detail in prior publications.   

In the autocatalysis-only model, a non-zero $ee$ is driven back to a racemic mixture. In this model, a continuous (though not necessarily constant) mechanism is required 
to drive the mixture towards homochirality.  In the original development of this model
\cite{farshid17}, it was assumed that stochastic mechanisms were responsible for
driving a mixture towards homochirality.  Here, we have replaced the stochastic
mechanism with a deterministic mechanism. In fact, the model presented here can be generalized to any deterministic mechanism.

It's also noted that in either model, if the non-autocatalytic rate $k_n = 0$, then the
enantiomeric excess is stable.  However, a zero non-autocatalytic rate will result in cumulative effects over time with no other influence.  

The MCA model, discussed in the introduction, could also explain both the enantiomeric excesses and the variations in isotopic abundance ratios, 
as a suggested site for it is a core-collapse supernova. It would also be subject to the processing of the electron anti-neutrinos. 
However, the MCA model does suffer from the lack of penetrability of the photons it needs to process the amino acids. The CPL model, however, requires a separate mode of processing to produce the isotopic abundance ratios.

Presented here is a rudimentary model in which the isotopic 
variations in meteoric amino acid are induced by external cosmic factors which affect
nuclei bound in amino acids. However, it should also be noted that nuclei not bound in
amino acids (those in free atoms or in external meteoric chemical constituents) are also
affected by the same external factors.  These external
nuclei may eventually contribute to the amino acids currently found in meteorites following recombination.  
They may affect the isotopic abundances in existing amino acids. However, becuase they
may be processed by the same mechanisms in this model which process
amino acids, it is unclear what the net effect would be. The nuclear effects changing isotopic abundances operate independently of the effects which select chirality.  Clearly, more
study is warranted in this context.

While this model has been applied to meteoritic amino acids, we note that other organic molecules have been found in 
meteorites~\cite[e.g., ][]{oba20,botta02,furukawa19,callahan11,glavin12,ehrenfreund01b} and 
comets~\cite{altwegg16}, 
and the search for
extraterrestrial organic molecules remains an active field in observational astronomy~\cite[e.g., ][]{mcguire18,sewilo19,ehrenfreund00,marel14} and planetary science~\cite[e.g., ][]{roos05,eigenbrode18,webster18,walsh14,sandford20}.	 However, in addition to chiral amino acids, other chiral organic molecules have been
found in meteorites with non-zero enantiomeric excesses~\cite{cooper16,glavin20} as well as isotopic ratios which differ from terrestrial values~\cite{glavin20}.  
Here the SNAAP model can be generalized and applied to other chiral molecules including polyols and sugars, both of which have been found in meteorites, with polyols
found to have non-zero $\delta^{13}C$~\cite{glavin20}.  Future work will study possible effects of the SNAAP model on other meteoritic organic molecules.

\acknowledgments
M.A.F. is supported by Moore
Foundation grant \#7799, a Western Michigan University Faculty Research and Creative Activities Award, and National Science Foundation Grant PHY1712832. 
T.K. is supported by Grants-in-Aid for Scientific Research of JSPS (15H03665, 17K05459).  T.O. is supported by a Grant-in-Aid for Scientific Research
of JSPS (18K03691). All authors
acknowledge support from the National Astronomical Observatory of Japan visiting professor program. The authors would also like to thank the 
referee for very insightful and encouraging comments.
\bibliography{references}

\end{document}